\documentclass[prd,twocolumn,preprintnumbers,superscriptaddress,nofootinbib]{revtex4}
\usepackage[a4paper, hdivide={1.91cm,,1.165cm}, vdivide={1.83cm,,3.6cm}]{geometry}

\usepackage{amstext,amssymb}
\usepackage{amsmath}
\usepackage{graphicx}
\usepackage[hyperfootnotes=true]{hyperref}
\usepackage{xspace}
\usepackage{color}
\usepackage{slashed} 

\begin{document}

\title{$0\nu\beta\beta$ in left-right theories with Higgs doublets and gauge coupling 
unification}
\author{Chayan Majumdar$^{1}$,  Sudhanwa Patra$^{2}$, Supriya Senapati$^{1}$, Urjit A. Yajnik}
\email{\\chayan@phy.iitb.ac.in \\
sudhanwa@iitbhilai.ac.in \\
supriya@phy.iitb.ac.in \\
yajnik@phy.iitb.ac.in}
\affiliation{Department of Physics, Indian Institute of Technology Bombay, Powai, Mumbai-400076 \\
$^{2}$Indian Institute of Technology Bhilai, GEC Campus, Sejbahar, Raipur-492015, Chhattisgarh, India}
\begin{abstract}
\vspace*{0.5cm}
We consider a version of Left-Right Symmetric Model in which the scalar sector consists of a Higgs bidoublet ($\Phi$) with $B-L=0$, 
Higgs doublets ($H_{L,R}$) with $B-L=1$ and a charged scalar ($\delta^+$) with $B-L=2$ leading to radiatively generated Majorana masses 
for neutrinos and thereby, leads to new physics contributions to neutrinoless double beta decay ($0\nu \beta \beta$). 
We show that such a novel framework can be embedded in a non-SUSY 
$SO(10)$ GUT leading to successful gauge coupling unification at around $10^{16}$~GeV with the scale of left-right symmetry breaking around 
$10^{10}$~GeV.
The model can also be extended to have left-right symmetry breaking at TeV scale, enabling detection of $W_R, Z_R$ bosons in LHC and future collider searches. 
In the context of neutrinoless 
double beta decay, this model can saturate 
the present bound from GERDA and KamLAND-Zen experiments. Also, we briefly explain how keV-MeV range RH neutrino arising from our model can saturate various astrophysical and cosmological constraints and can be considered as warm Dark Matter (DM) candidate to address various cosmological issues. We also discuss on left-right theories with Higgs doublets without having scalar bidoublet leading to 
fermion masses and mixings by inclusion of vector like fermions.
\end{abstract}

\maketitle

\section{Introduction}
\label{sec:intro}
The Standard Model (SM) 
is a remarkbaly successful theory for Particle Physics in accord with almost all data till current accelerator reach.
However several open problems persist which cannot be addressed within SM. One such problem is the parity asymmetry seen in low-energy weak-interactions while the strong interactions are parity-conserving. 
It is believed that SM can be thought of as the effective low energy theory of a larger framework which is parity symmetric at higher energy scale. 
From recent neutrino oscillation experiments~\cite{Fukuda:2001nk, Ahmad:2002jz}, 
there is convincing evidence for neutrino masses; which are not permitted in the SM. 
Within the framework of the left-right symmetric models (LRSM)~\cite{Mohapatra:1974gc, Pati:1974yy, Senjanovic:1975rk,Senjanovic:1978ev,
Mohapatra:1979ia,Mohapatra:1980yp,Pati:1973uk,Pati:1974vw} , 
we can have possible resolutions for both the problems. In this framework, the fundamental interactions are parity-even at energy scales much above the electroweak scale. 
Such a scenario naturally admits right-handed neutrinos with non-zero masses.

In this work, we consider a version of LRSM in which the scalar sector consists of a Higgs bidoublet ($\Phi$) with $B-L=0$, 
Higgs doublets ($H_{L,R}$) with $B-L=1$. 
With this particle content, quarks and leptons including neutrinos can obtain Dirac masses.
The manifest left-right symmetric models with Higgs triplets and bidoublet
\cite{Mohapatra:1979ia}
provide Majorana masses to neutrinos and hence, allow lepton number violation. However, within this version of LRSM with Higgs doublets and bidoublet, 
there are no Majorana mass terms and thus, no lepton number violation in the theory.  In order to have lepton number violation or Majorana 
mass terms~\cite{Keung:1983uu, Ferrari:2000sp, Nemevsek:2011hz, Nemevsek:2018bbt}, the model is expanded by adding a charged scalar $\delta^+$ with $B-L$ charge $2$ which will allow us to generate 
the Majorana mass terms for neutrinos at loop-level as first pointed out by P. Fileviez Perez et al.~\cite{FileviezPerez:2016erl}. We can consider this version of LRSM as Fileviez Perez-Murgui-Ohmer (FPMO) model. In this reference, they also have discussed the collider signatures of the Lepton Number Violating (LNV) processes in the context of this left-right symmetric model.  

Neutrinoless double beta decay ($0\nu\beta\beta$) is a decay mode of a given isotope in which two neutrons simultaneously convert 
into two protons and two electrons without being accompanied by any neutrinos. The experimental observation of such 
a rare process would reveal the Majorana nature of light neutrinos~\cite{Majorana:1937vz} 
indicating the violation of Lepton Number and can provide information on the absolute scale of neutrino mass. Till date, the best 
lower limit on half-life of the neutrinoless double beta decay using $^{76}\mbox{Ge}$ is $T^{0\nu}_{1/2}\, > 
8.0 \times 10^{25}$ yrs. at 90\% C.L. from GERDA~\cite{Agostini:2013mzu}. For $^{136}\mbox{Xe}$ isotope, the derived 
lower limits on half-life from KamLAND-Zen experiment is $T^{0\nu}_{1/2}\, > 1.6 \times 10^{26}$ yrs 
\cite{Gando:2012zm}. The proposed sensitivity of the future planned nEXO experiment is $T_{1/2}^{0\nu}(^{136}\text{Xe}) 
\approx 6.6\times 10^{27}$~yrs \cite{Albert:2014afa}.

The Lepton number violating $0\nu\beta\beta$ process could arise either from the standard mechanism due to exchange 
of light Majorana neutrinos or by some new physics beyond SM (BSM). 
The manifest LRSM provides us the existence of right-handed neutrinos, light neutrino masses, new right-handed massive gauge bosons and their mixing with the left-handed counterpart gauge bosons and the possibility of light-heavy neutrino mixing~\cite{Mohapatra:1980yp, Mohapatra:1981pm, 
Hirsch:1996qw, Tello:2010am, Chakrabortty:2012mh, Patra:2012ur, Awasthi:2013ff, Barry:2013xxa, Dev:2013vxa, Ge:2015yqa, 
Awasthi:2015ota,Halprin:1983ez}. In the present scenario, we aim to discuss new physics contributions to neutrinoless 
double beta decay within a version of left-right symmetric model with Higgs doublets and bidoublet where Majorana masses for 
neutrinos are generated at loop-level. We also intend to examine the resulting contributions to $0\nu\beta\beta$ 
transition which can saturate the current experimental bounds. 

Grand Unified Theories (GUTs)~\cite{Pati:1974yy, Georgi:1974sy, Georgi:1974yf, Fritzsch:1974nn} based on the gauge group $SO(10)$ are very appealing in which the three fundamental 
forces strong, weak and electromagnetic have a common origin. They have potential to shed light on many unsolved questions of SM.  Unlike the $SU(5)$ GUT which breaks directly to SM, $SO(10)$ admits 
intermediate symmetry breaking like left-right symmetry or Pati-Salam symmetry. Our goal here is also to embed the left-right symmetric theory into such a non-supersymmetric $SO(10)$ GUT. Such left-right symmetry breaking occuring at the scale of a few TeV can give rise to 
interesting weak phenomenology i.e, right-handed gauge bosons $W_R, Z_R$ at collider scales. 

The structure of the paper is as follows. Section \ref{sec:mod} contains a brief introduction of the FPMO model including the particle content and the symmetry breaking pattern. The generation of Dirac and Majorana masses and the resulting neutral lepton mass matrix have been discussed in section \ref{sec:mass}. In the subsequent sections \ref{sec:GUT} and \ref{Gauge}, we 
embed this LRSM version in a non-SUSY $SO(10)$ GUT framework. In section \ref{sec:beta}, we discuss the new physics contributions to neutrinoless double beta decay which can saturate the KamLAND-Zen and GERDA experiments. Also, in section \ref{cosmo}, we briefly explain the validity of our model in various cosmological scenario. We 
discuss fermion masses and mixings in left-right symmetric models with Higgs doublets and without having scalar bidoublet in section \ref{sec:lrsm-universal}. 
In Appendix \ref{sec:a1}, we present the full Lagrangian of the framework and the minimization of the scalar potential has been carried out in Appendix \ref{sec:a2}. 

\section{\hspace*{-0.4cm} Description of the model}
\label{sec:mod}
The left-right symmetric model~\cite{Mohapatra:1974gc, Pati:1974yy, Senjanovic:1975rk,Senjanovic:1978ev,
Mohapatra:1979ia,Mohapatra:1980yp,Pati:1973uk,Pati:1974vw} is based on the gauge group, 
\begin{align}
\mathcal{G}_{LR}\equiv SU(2)_L \times SU(2)_R \times U(1)_{B-L} \times SU(3)_{C}\, 
\label{eq:LRSM}
\end{align}
where the electric charge is defined as 
\begin{align}
Q=T_{3L}+T_{3R}+\frac{B-L}{2}\, .
\end{align}
Under this left-right symmetric gauge group, the usual quarks and leptons transform as
\begin{eqnarray}
&&q_{L}=\begin{pmatrix}u_{L}\\
d_{L}\end{pmatrix}\equiv[2,1,1/3,3]\,, ~ q_{R}=\begin{pmatrix}u_{R}\\
d_{R}\end{pmatrix}\equiv[1,2,1/3,3]\,,\nonumber \\
&&\ell_{L}=\begin{pmatrix}\nu_{L}\\
e_{L}\end{pmatrix}\equiv[2,1,-1,1] \, , ~ \ell_{R}=\begin{pmatrix}\nu_{R}\\
e_{R}\end{pmatrix}\equiv[1,2,-1,1] \, . \nonumber
\end{eqnarray}

The left-right symmetric model can be spontaneously broken down to SM gauge group $SU(2)_L\times U(1)_Y \times SU(3)_C$ 
either using Higgs doublets or Higgs triplets or combination of both having non-zero $B-L$ charges. In manifest left-right symmetric models with Higgs 
triplets, the model accomodates lepton number violation via Majorana masses for left-handed and right-handed neutrinos at 
tree level through non-zero VEVs of these triplets. In the present framework Higgs doublet $H_R$ breaks the left-right symmetry to SM. 
The left-right symmetry demands the existence of Higgs doublet $H_L$ which is the left counterpart of $H_R$.  We also need a 
Higgs bidoublet $\Phi$ with $B-L=0$ to break the SM electroweak gauge group $SU(2)_L\times U(1)_Y \times SU(3)_C$ down to 
$U(1)_{Q} \times SU(3)_C$.  Thus, the symmetry breaking pattern for this left-right symmetric model 
is given by
\begin{align}
\mathcal{G}_{LR} 
\mathop{\longrightarrow}^{\langle H_R \rangle} \mathcal{G}_{SM} {\large \mathop{\longrightarrow}^{\langle H_L\rangle, \langle \Phi \rangle}} U(1)_{Q} \times SU(3)_C
\end{align}
where
\begin{align}
&\mathcal{G}_{LR} \equiv SU(2)_L \times SU(2)_R \times U(1)_{B-L} \times SU(3)_C\, , \nonumber \\
&\mathcal{G}_{SM} \equiv SU(2)_L \times U(1)_{Y} \times SU(3)_C\, , \nonumber
\end{align}
In this left-right symmetric model with Higgs doublets and bidoublets, all the fermions including 
neutrinos are getting Dirac type masses and thus, have no lepton number violation in the model. The lepton number 
violation can be incorporated minimally with the inclusion of a charged scalar $\delta^+(1_L,1_R,2_{B-L},1_C)$. 
We shall discuss in the next section how Majorana masses for both left-handed and right-handed neutrinos are generated 
at one-loop level with the help of this extra charged scalar. Thus, the complete scalar sector of the model is 
given by~\cite{FileviezPerez:2016erl}
\begin{eqnarray}
&&
\Phi=
\begin{pmatrix} 
\phi_{1}^0     &  \phi_{2}^+ \\
\phi_{1}^-     &  \phi_{2}^0
\end{pmatrix} \equiv[2,2,0,1]\,,\nonumber \\
&&
H_L=
\begin{pmatrix} 
h_L^+\\
h_L^0
\end{pmatrix}\equiv[2,1,1,1]  \,,\nonumber \\
&&
H_R=
\begin{pmatrix} 
h_R^+  \\
h_R^0
\end{pmatrix}\equiv[1,2,1,1] \,,\nonumber \\
&&
\delta^+ \equiv [1,1,2,1]
\end{eqnarray}

\section{Neutrino Masses}
\label{sec:mass}
The leptonic Yukawa interaction Lagrangian can be read as
\begin{eqnarray}
\mathcal{L}_{\rm Yuk}&=&\overline{q_L}  \left(Y^q\, \Phi  + \widetilde{Y^q} \widetilde{\Phi} \right) \, q_{R}\nonumber \\
     &&+\overline{\ell_{L}}  \left(Y^\ell\, \Phi  + \widetilde{Y^\ell} \widetilde{\Phi} \right) \, \ell_{R}\nonumber \\
     &&+\lambda^L \ell^T_L C \ell_L \delta^+ + \lambda^R \ell^T_R C \ell_R \, \delta^+ + \mbox{h.c.}\,
\end{eqnarray}
The VEVs of the Higgs scalars are taken to be, 
\begin{eqnarray}
&&\langle \Phi \rangle = \begin{pmatrix} v_1  & 0 \\  0  & v_2 \end{pmatrix}\, , \quad \langle \delta^+ \rangle =0\, ,  \nonumber \\
&&\langle H_R \rangle = \begin{pmatrix} 0 \\  v_R \end{pmatrix}\, , \quad 
\langle H_L \rangle = \begin{pmatrix} 0 \\  v_L \end{pmatrix}\, . \quad
\label{vev} 
\end{eqnarray}

\begin{figure}[t!]
\includegraphics[width=0.99\linewidth]{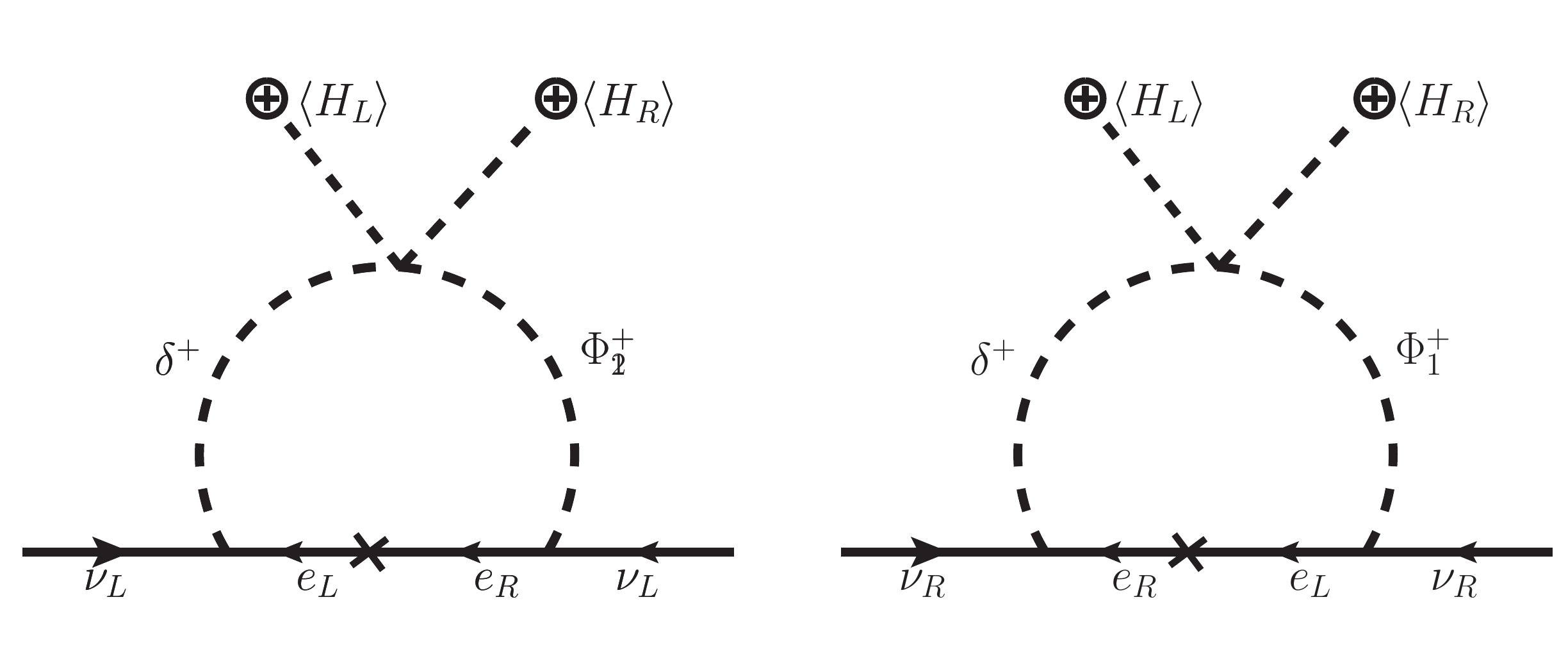}
\caption{Generation of Majorana masses for left-handed and right-handed neutrinos at one loop level.}
\label{feyn:seesaw}
\end{figure}
After spontaneous symmetry breaking the quarks, charged leptons and neutrinos get their Dirac type masses as,
\begin{eqnarray}
&&  M_u =  Y^q\, v_1 + \widetilde{Y^q} v^*_2\,, \quad \quad 
    M_d =  Y^q\, v_2 + \widetilde{Y^q} v^*_1\,, \quad \quad \nonumber \\
&& M_D = Y^\ell\, v_1 + \widetilde{Y^\ell}\, v^*_2 \,, \quad \quad 
    M_e =  Y^\ell\, v_2 + \widetilde{Y^\ell}\, v^*_1 \,.
\end{eqnarray}
It should be noted that the electroweak VEV $v_{EW}$ can be expressed as,
\begin{eqnarray}
v^2_{EW} = v^2_1 + v^2_2 + v^2_{L}
\end{eqnarray}
It is possible that one of the VEVs of $\Phi$ can be chosen to be small. 
In the limit, $v_2 \rightarrow 0$~(\cite{FileviezPerez:2016erl}),
the fermion masses 
are given by
\begin{eqnarray}
&&  M_u =  Y^q\, v_1 \,, \quad \quad 
    M_d =  \widetilde{Y^q} v^*_1\,, \quad \quad \nonumber \\
&& M_D = Y^\ell\, v_1  \,, \quad \quad 
    M_e =   \widetilde{Y^\ell}\, v^*_1 \,.
    \label{elec}
\end{eqnarray}
Thus, one can write down the up-type and down-type quark masses from $v_1$ as,
\begin{eqnarray}
&&  M_u =  Y^q\, v_1 =  V^u_L \begin{pmatrix}  m_u & 0 & 0 \\ 0 & m_c & 0 \\  0 & 0 & m_t  \end{pmatrix}  {V^u_R}^\dagger\, \nonumber \\
 &&  M_d =  \widetilde{Y^q} v^*_1= V^d_L \begin{pmatrix}  m_d & 0 & 0 \\ 0 & m_s & 0 \\  0 & 0 & m_b  \end{pmatrix}  {V^d_R}^\dagger\, 
\end{eqnarray}
leading to CKM mixing matrices as,
\begin{eqnarray}
V_{\rm CKM} = {V^u_L}^\dagger V^d_L\, \quad, \quad V_{\rm R} = {V^u_R}^\dagger V^d_R\,.
\end{eqnarray}
For simplicity, we can work in basis where down-type quark masses are already diagonal i.e, $\widetilde{Y^q}$ as diagonal matrix and other Yukawa 
matrices can be constructed by the physical up and down-type quark masses along with CKM mixing matrix.

Before commenting on $\widetilde{Y^\ell}$ and $Y^\ell$, let us discuss the one-loop generated Majorana masses for left-handed and right-handed 
neutrinos (pointed out in ref.~\cite{FileviezPerez:2016erl}) as shown in Fig.\ref{feyn:seesaw} as,
\begin{eqnarray}
M_{L}^{\text{1-loop}} \simeq  \frac{\lambda^\prime \langle H_L \rangle \langle H_R \rangle}{16 \pi^2}  \frac{\lambda^L M_\ell Y^T_\ell}{M^2} \mathcal{I}\, ,   \nonumber \\
M_{R}^{\text{1-loop}} \simeq  \frac{\lambda^\prime \langle H_L \rangle \langle H_R \rangle}{16 \pi^2}  \frac{\lambda^R M_\ell Y^T_\ell}{M^2} \mathcal{I}\, , 
\label{eqn:loop}  
\end{eqnarray}
where $M={\it max}(M_{\delta^+}, M_\Phi)$, $M_\ell$ is the mass of the lepton and $\mathcal{I}$ is the loop factor, can be found as,
\begin{center}
$\mathcal{I}=\frac{log[\frac{M_{\ell}^{2}}{M_{\delta ^{+}}^{2}}]M_{\delta ^{+}}^{2}}{M_{\delta ^{+}}^{2}-M_{\ell}^{2}}-\frac{log[\frac{M_{\ell}^{2}}{M_{\Phi}^{2}}]M_{\Phi}^{2}}{M_{\Phi}^{2}-M_{\ell}^{2}}$
\end{center}
  Thus, the complete neutral lepton mass matrix is
\begin{eqnarray}
\mbox{M}= \begin{pmatrix}  M_{L}^{\text{1-loop}}  & M_D \\ M^T_D  & M_{R}^{\text{1-loop}}  \end{pmatrix}
\end{eqnarray}
In the mass hierarchy $M_{R}^{\text{1-loop}} \gg M_D \gg M_{L}^{\text{1-loop}}$ the light and heavy neutrino masses using seesaw approximation and in the 
limit $M_{L}^{\text{1-loop}} \to 0$ as,
\begin{eqnarray}
m_\nu = - M_D (M_{R}^{\text{1-loop}})^{-1} M^T_D  \, , \quad m_R = M_{R}^{\text{1-loop}}\,.
\end{eqnarray}

\section{Embedding the framework in $SO(10)$ GUT}
\label{sec:GUT}
We embed this version of left-right symmetric model for lepton number violation as discussed in Section~\ref{sec:mod} within a non-supersymmetric $SO(10)$ GUT 
to predict the scale of left-right symmetry breaking scale~\cite{Bhatt:2008dg,Fukuyama:2004ps}. The symmetry breaking chain of $SO(10)$ GUT is 
      \begin{align}
      \label{eq:BreakingChain}
      & SO(10)\, \mathop{\longrightarrow}^{M_U} \mathcal{G}_{2_L 2_R 1_{B-L} 3_C}\mathop{\longrightarrow}^{M_{R}} \mathcal{G}_{2_L 1_Y 3_C}\, \big(\mbox{SM}\big)
     \mathop{\longrightarrow}^{M_{Z}} \mathcal{G}_{1_{Q} 3_C}\,. \nonumber 
      \end{align}
The $SO(10)$ breaks down to the SM gauge group with the intermediate breaking step $\mathcal{G}_{2_L 2_R 1_{B-L} 3_C}$ at $M_{R}$ scale. 
At the first stage the symmetry breaking for $SO(10)$ GUT to the left-right gauge group $\mathcal{G}_{2_L 1_R 1_{B-L} 3_C}$ at unification scale 
$M_U$ is achieved by assigning a non-zero vev to a Higgs field $\langle \Sigma(1, 1, 15) \rangle \in \{45_H\}$.  The subsequent stage of symmetry 
breaking of $\mathcal{G}_{2_L 2_R 1_{B-L} 3_C} \to \mathcal{G}_{SM}$ is done by assigning a  non-zero VEV to Higgs doublet $H_R(1, 2,1,1) 
\in 16_H$ with $B-L=1$.

For the generation of the SM fermion masses the Higgs multiplets are limited as 16 $\times$ 16 = $ 10_{s} +120_{a} + \overline{126}_{s} $. The Higgs field $ H_{\Phi} \equiv h $ belonging to 10-dimensional representation of $SO(10)$ decomposes under left-right gauge group as
\begin{center}
$ H_{\Phi}(10) = \Phi(1,2,2,0) \oplus (3,1,1, -\frac{1}{3}) \oplus (\overline{3},1,1, \frac{1}{3}) $.
\end{center}
So, clearly we can see that the bi-doublet $ \Phi(1,2,2,0) $ in the left-right model belongs to $ H_{\Phi}(10) $. Also, the new Higgs field $\delta^+$ belongs to 120-dimensional representation of $SO(10)$. 

For a $SO(10)$ GUT model the fermion and gauge sector are much simpler than the Higgs sector because it is required both for generating fermion masses as well as
the breaking of the $SO(10)$ gauge group down to the SM gauge group. First of all, to break the $SO(10)$ gauge group to the left-right gauge group, one needs Higgs field either $A(210)$ or $ B(45)$. The deomposition of these fields under Pati-Salam group ($SU(2)_{L} \otimes SU(2)_R \otimes SU(4)_C $) are as follows,
\begin{center}
$ A(210) = (1,1,1) \oplus (1,1,15) \oplus (2,2,6) \oplus (3, 1, 15) \oplus (1,3, 15) \oplus (2,2, 10) \oplus (2, 2, \overline{10}) $ \\
$ B(45) = (3,1,1) \oplus (1,3,1) \oplus (1,1, 15) \oplus (2, 2, 6) $
\end{center}

To have the breaking of the gauge group we can give \textit{vev} to any one of these fields in the singlet direction. In our model D-parity is conserved until left-right group is broken. As 210-dimensional Higgs representation is D-parity odd then to break the $SO(10)$ gauge group to the left-right gauge group we will use here $B(45)$. 

Now, we have to embed Higgs doublets $H_L (2,1,1,1)$ and $H_R (1,2,1,1)$ in some tensor representation of $SO(10)$. From the quantum numbers we can embed them in the spinorial Higgs representation ($C(16) \oplus \overline{C(16)}$). The $16 \oplus \overline{16}$ spinor representation decomposes under the left-right symmetric group as
\begin{align}
C(16) &= H^{\ast}_L (2,1,-1,1) \oplus H_R (1,2,1,1) \nonumber \\
& \oplus (2,1,\frac{1}{3},3) \oplus (1,2,-\frac{1}{3},\overline{3}) \nonumber
\end{align}
\begin{align}
\overline{C(16)} &= H_L (2,1,1,1) \oplus H^{\ast}_R (1,2,-1,1) \nonumber \\
&\oplus (1,2,\frac{1}{3},3) \oplus (2,1,-\frac{1}{3},\overline{3}) \nonumber
\end{align}

Having embedded all the Higgs fields of our model into $SO(10)$ tensor field to remain SM intact at left-right breaking scale we can choose the \textit{vev} along the corresponding singlet directions of $B(45)$ as
\begin{center}
$\langle B \rangle = M_U \hat{B}_U + M_R \hat{B}_R$
\end{center}
with $\hat{B}_U$ and $\hat{B}_R$ defining the singlet directions under the SM gauge group given as
\begin{center}
$\hat{B}_U = (\hat{B}_{56} + \hat{B}_{78} + \hat{B}_{910})$ \\
$\hat{B}_{R} = (\hat{B}_{12} + \hat{B}_{34})$
\end{center}
Here the indices $(1, 2, 3, 4)$ and $(5, 6, 7, 8, 9, 10)$ belong to $SO(4)$ and $SO(6)$ respectively which are the subgroups of $SO(10)$ . We are assuming that the normalization factors are absorbed in the corresponding \textit{vev} values.

So, the most general $SO(10)$ invariant Higgs potential can be written as
\begin{align}
V &= \mu _{B}^2 B_{ab}B_{ba} + \mu _{h}^2 h_a h_a + \mu _{C}^2 (\overline{C}C) + \mu _{\delta}^2 (\delta ^+)^2 \nonumber \\
&+ \lambda _{\delta} (\delta ^+)^4 + \lambda _{B} B^2 B^2 + \lambda ^{\prime} _{B} B^4 + \lambda _{h} h^4 + \lambda _{C} (\overline{C}C)^2 \nonumber \\
&+ \lambda ^{\prime} _{C} (C^4 + \overline{C}^4) + h_a g_{hB} B_{ab} B_{bc} h_c + g^{\prime}_{hB} B^2 h^2 \nonumber \\
&+ (g_{hC} h^2 + g_{BC} B^2) \overline{C}C + g_{\delta h C} (\delta ^+)^2 h (CC + \overline{C}\overline{C})
\end{align}
Now, assigning \textit{vev} to the field $B(45)$ in the potential as
\begin{center}
$\langle B \rangle = i\tau _2 \otimes diag (M_R, M_R, M_U, M_U, M_U)$
\end{center}
(here we have just replaced the vevs with the corresponding breaking scales for convinenience) and with $\langle \delta^{+} \rangle $, the potential can be written as in terms of \textit{vev}s as
\begin{align}
V_{\text{para}} &= \mu _{B}^2 (6M_{U}^2 + 4M_{R}^2) + \mu _{h}^2 h_a h_a + \mu _{C}^2 (\overline{C}C) \nonumber \\
&+ \lambda _{B} (6M_{U}^2 + 4M_{R}^2)^2 + \lambda ^{\prime} _{B} (6M_{U}^4 + 4M_{R}^4) + \lambda _{h} h^4 \nonumber \\
&+ \lambda _{C} (\overline{C}C)^2 + \lambda ^{\prime} _{C} (C^4 + \overline{C}^4) + g_{hB} 4M_{R}^2 h_a h_a (a = \nonumber \\
& 1,2,6,7) + g_{hB} 6M_{U}^2 h_a h_a (a = 3,4,5,8,9,10)\nonumber \\
& + g^{\prime}_{hB} (6M_{U}^2 + 4M_{R}^2) h^2  + g_{hC} h^2 \overline{C}C \nonumber \\
&+ g_{BC} (6M_{U}^2 + 4M_{R}^2)\overline{C}C
\end{align}

To accomodate our model in recent collider phenomenology, we have to bring down the left-right breaking at around some $TeV$ scale, we have to extend our model by introducing a new particle $\xi(1,1,4/3,6)$ which belongs to 54-dimensional representation $D(54)$ of $SO(10)$. Under Pati-Salam gauge group $D(54)$ decomposes as
\begin{center}
$D(54) = (1,1,1) \oplus (3,3,1) \oplus (1,1,20) \oplus (2,2,6)$
\end{center}

We can choose \textit{vev} $\langle D \rangle=M_U \hat{D}$ with 
\begin{center}
$\hat{D} = 3\sum _{a=1}^{4}\hat{D}_{aa} - 2\sum _{a=5}^{10}\hat{D}_{aa}$
\end{center}
where we take the singlet direction to keep SM intact at LRSM breaking scale.

Now including $\xi$ particle we can write the $SO(10)$ invariant Higgs potential as
\begin{align}
V^{\prime} &= V + \mu _{D}^2 D_{ab} D_{ba} + \lambda _{D}D^4 + g_{BD}B^2 D^2 + \nonumber \\
&g^{\prime}_{BD} B_{ab} B_{bc} D_{cd} D_{da} + g^{\prime \prime}_{BD} B_{ab} D_{bc} B_{cd} D_{da} \nonumber \\
& + h_a g_{hD} D_{ab} D_{bc} h_c + g_{DC} D^2 \overline{C}C 
\end{align}

In this extended Higgs sector potential, we can put \textit{vev} for $D(54)$ as,
\begin{center}
$\langle D \rangle = \textit{I} \otimes diag(-\frac{3}{2}M_U, -\frac{3}{2}M_U, M_U, M_U, M_U)$
\end{center}
So, the parametrized potential can be rewritten as,
\begin{align}
V^{\prime}_{\text{para}} &= V_{\text{para}} + \mu _{D}^2 15M_U^2 + \lambda _{D}\frac{105}{4}M_U^4 + g_{BD}M_U^2(6M_U^2 + 9M_R^2) \nonumber \\
& + g_{hD}[9M_U^2 h_a h_a (a = 1,2,6,7) + 4M_U^2 h_a h_a (a = 3,4,5,\nonumber \\
& 8,9,10)] + g_{DC} 15M_U^2 \overline{C}C 
\end{align}
Here we have considered $M_R \ll M_U$ while writing the potential.

\section{Gauge Coupling Evolution}
\label{Gauge}
The relevant one-loop RG equation~\cite{Jones:1981we} for the gauge couplings $g_i$ ($i=2L, Y, 3C$) from SM to LRSM and $g_i$ ($i=2L, 2R, BL, 3C$) from LRSM 
to GUT scale,
\begin{equation}
\mu\,\frac{\partial g_{i}}{\partial \mu}=\frac{b_i}{16 \pi^2} g^{3}_{i},
\label{eq:RG}
\end{equation}
where the one-loop beta-coefficients $b_i$ are as follows,
\begin{eqnarray}
	&&b_i= - \frac{11}{3} \mathcal{C}_{2}(G) 
				 + \frac{2}{3} \,\sum_{R_f} T(R_f) \prod_{j \neq i} d_j(R_f) \nonumber \\
  &&\hspace*{2.5cm} + \frac{1}{3} \sum_{R_s} T(R_s) \prod_{j \neq i} d_j(R_s).
\label{oneloop_bi}
\end{eqnarray}
In the above formula, $\mathcal{C}_2(G)$ is the quadratic Casimir operator for gauge bosons in their adjoint representation,
\begin{equation}
	\mathcal{C}_2(G) \equiv 
	\begin{cases} 
		N & \text{if } SU(N), \\
    0 & \text{if }  U(1).
	\end{cases}
\end{equation}
$T(R_{f,s})$ are the traces of the irreducible representation $R_{f,s}$ for a given fermion (scalar),
\begin{equation}
	T(R_{f,s}) \equiv 
	\begin{cases} 
		1/2 & \text{if } R_{f,s} \text{ is fundamental}, \\
    N   & \text{if } R_{f,s} \text{ is adjoint}, \\
		0   & \text{if } U(1).
	\end{cases}
\end{equation}
and $d(R_{f,s})$ is the dimension of a given representation $R_{f,s}$ under all $SU(N)$ gauge groups 
except the $i$-th~gauge group under consideration. 

\begin{figure}[t!]
\includegraphics[width=0.95\linewidth]{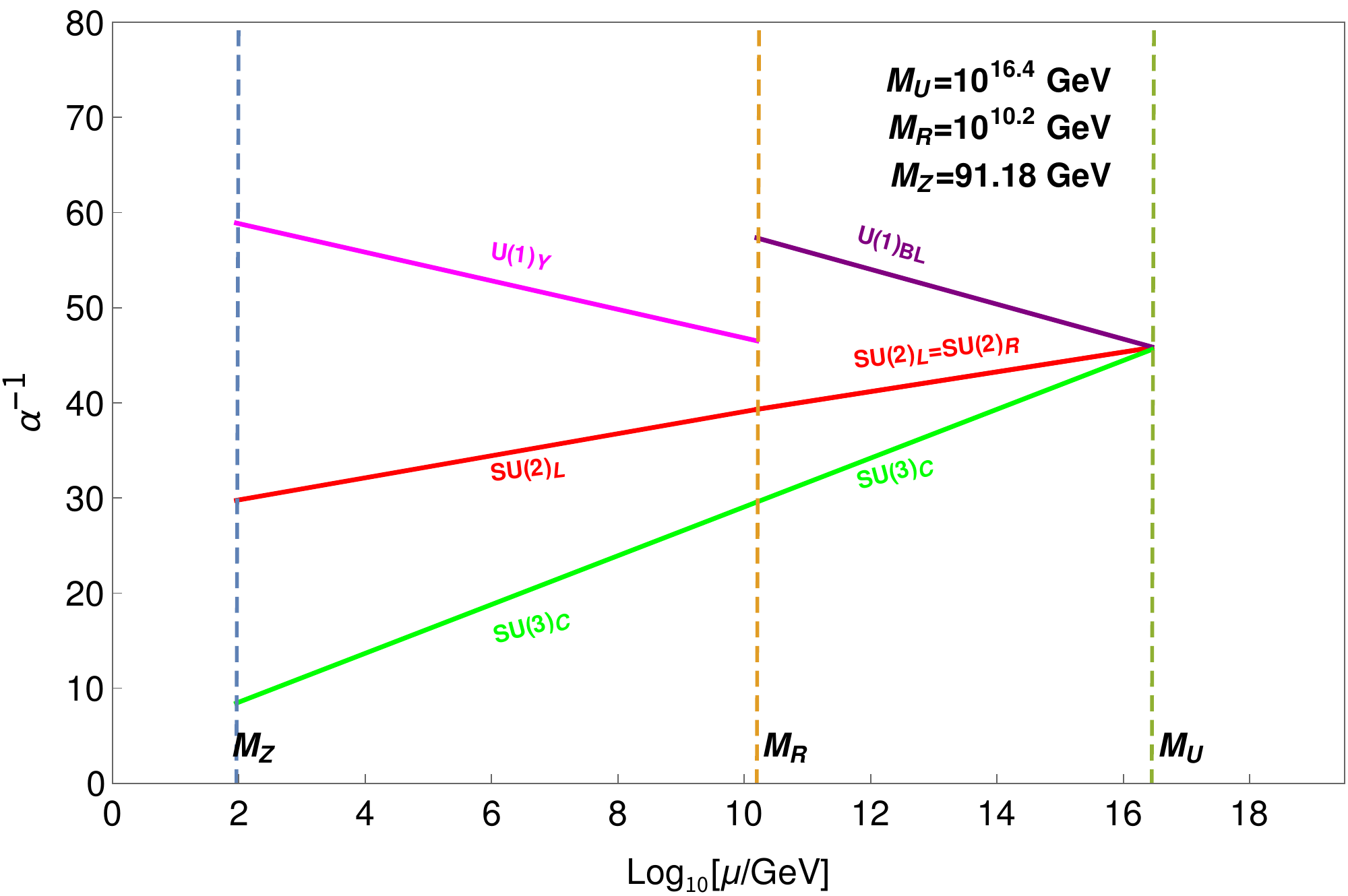}
\caption{Gauge coupling running and unification in case 1, this version of left-right symmetric model 
considered here. It implies unification scale $M_U = 10^{16.4}$~GeV and left-right 
breaking scale $M_\text{R} = 10^{10.2}$~GeV.}
\label{fig:unifn}
\end{figure}

\begin{figure}[t!]
\includegraphics[width=0.95\linewidth]{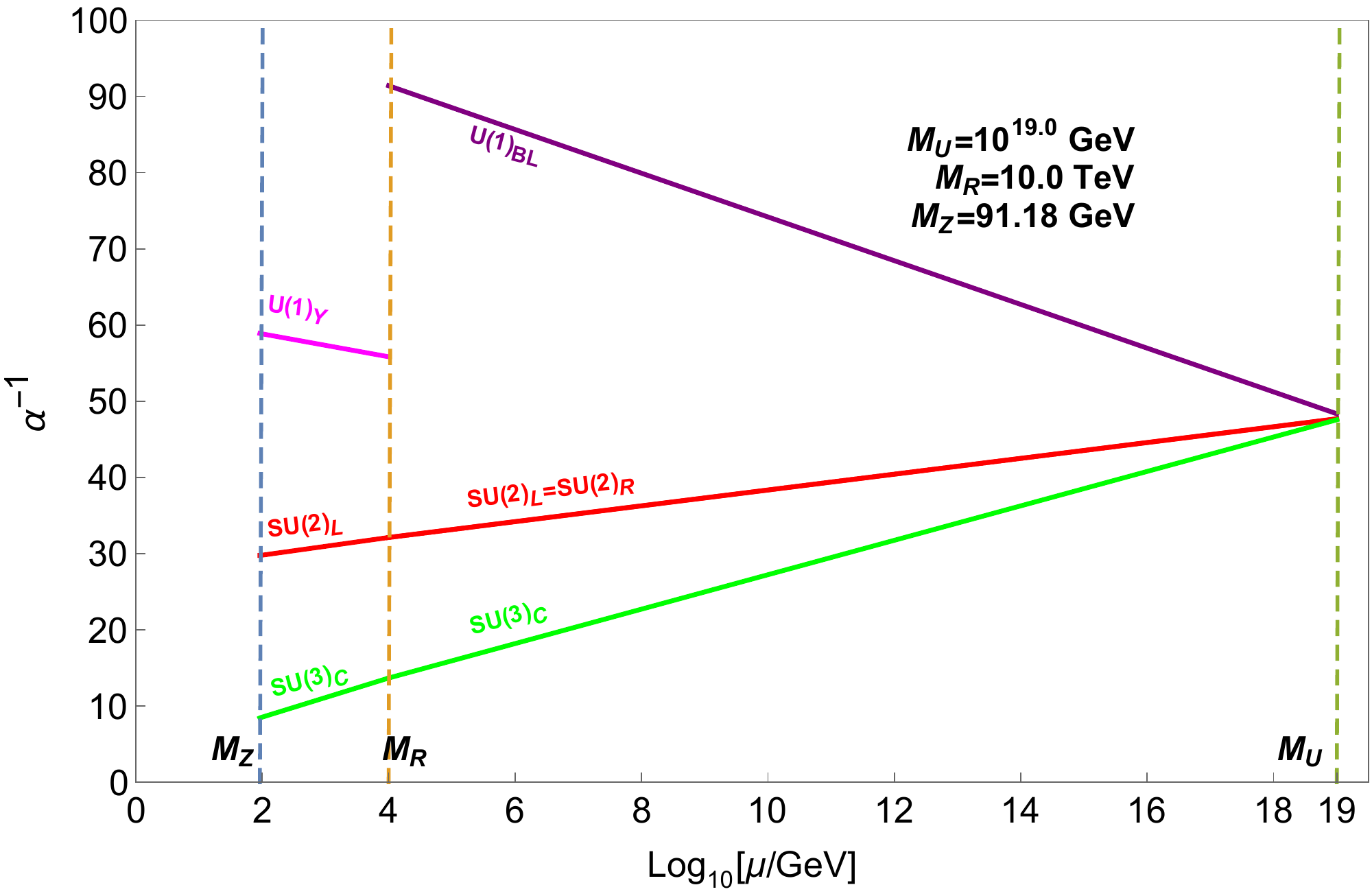}
\caption{Gauge coupling running and unification in case 2, with the model of Fig.\ref{fig:unifn} 
enhanced by addition of four copies of charged scalar $\delta^{+}(1_L, 1_R, 2_{BL}, 1_C)$ 
and one copy of $\xi(1_L, 1_R, 4/3_{BL}, 6_{C})$ at mass scale $M_R$ and above.
This results in $M_U$ being pushed close to Planck scale $10^{19}$GeV however the left-right 
symmetry breaking scale $M_\text{R}$ becomes $= 10^{4}$~GeV.
}
\label{fig:unifn-TeV}
\end{figure}

When we consider the unification of this model in $SO(10)$, denoted case 1, we find that it leads 
to a high value for $M_R$. Thus additionally we shall also consider a model that permits a 
scale for $M_R$ closer to the TeV scale, by introducing additional scalar multiplets as introuduced 
in case 2 below. Using the particle content of the model, the 
one-loop beta coefficients for different mass range are as follows,

\begin{eqnarray}
&&\hspace*{-0.4cm} {\bf \mbox{(i)}\, \mu \in [M_Z - M_{R}] }:  \nonumber \\
        & &\hspace*{0.5cm} G_{213} \equiv {\small SU(2)_L\times U(1)_{Y} \times SU(3)_C}, \nonumber \\
        & &\hspace*{0.5cm} \mbox{Higgs:\,}\phi (2,1/2,1) \subset 10_H\, ; \nonumber \\
        & &\hspace*{0.5cm} \pmb{b}_{2L}=-19/6,\,\pmb{b}_{Y}= 41/10,\, \pmb{b}_{3C}=-7\, , \nonumber \\
&&\hspace*{-0.4cm} {\bf \mbox{(ii)}\, \mu \in [M_{R}-M_U] }:  \nonumber \\
        & &\hspace*{0.5cm} G_{2213} \equiv {\small SU(2)_L\times SU(2)_R\times U(1)_{B-L} \times SU(3)_C}, \nonumber \\
        & & a) \text{Case 1:}\nonumber \\
        & &\hspace*{0.5cm} \mbox{Higgs:\,}\Phi (2,2,0,1), H_L(2,1,1,1), \, H_R (1,2,1,1), \nonumber 
\\
        & &\hspace*{1.5cm}  \, \delta^+(1,1,2,1) \nonumber \\ 
        & &\hspace*{0.5cm} \pmb{b}^{\prime}_{2L}=-17/6,\,\pmb{b}^{\prime}_{2R}=-17/6 ,\, 
                           \pmb{b}^{\prime}_{BL}=5,\, \pmb{b}^{\prime}_{3C}=-7\, . \nonumber \\
        & & b) \text{Case 2:}\nonumber \\
        & &\hspace*{0.5cm} \mbox{Higgs:\,}\Phi (2,2,0,1),  
        H_L(2,1,1,1), H_R (1,2,1,1),\nonumber \\
        & & \hspace*{0.5cm} \text{and new fields}, \nonumber \\
        & &\hspace*{1.5cm}\xi(1,1,4/3,6) \, \text{and 4 copies of }\,\delta^+(1,1,2,1) 
\nonumber\\ 
        & &\hspace*{0.5cm} \pmb{b}^{\prime}_{2L}=-17/6,\,\pmb{b}^{\prime}_{2R}=-17/6 ,\, 
                           \pmb{b}^{\prime}_{BL}=47/6,\nonumber \\
       & & \pmb{b}^{\prime}_{3C}=-37/6\, . \nonumber \\
\label{higgs-so10}
\end{eqnarray} 
The evolution of the gauge couplings in cases 1 and 2 are displayed in Fig. \ref{fig:unifn} and Fig.
\ref{fig:unifn-TeV} respectively.

The two unknown parameters,  left-right symmetry breaking scale $M_R \simeq \langle 
H_R \rangle$ 
and the unification scale $M_U$ can be solved for by considering the 
RG equations for individual gauge couplings and extrapolating them to $M_U$. 
Using Eq. (\ref{eq:RG}), 
the key equations are
\begin{eqnarray}
&&\mathcal{A}_U \mbox{ln}\left( \frac{ {M_U} }{M_Z}\right)            
        +\mathcal{A}_R \mbox{ln}\left(\frac{ {M_R} }{M_Z}\right) 
        =\mathcal{C}_1
\label{gLgR:rel1}\, \\
&&\mathcal{B}_U \mbox{ln}\left( \frac{M_U}{M_Z}\right)            
        +\mathcal{B}_R \mbox{ln}\left( \frac{M_R}{M_Z}\right) 
        =\mathcal{C}_2
\label{gLgR:rel2}\,
\end{eqnarray}
Here
\begin{eqnarray}
&&\mathcal{C}_1=16\pi\left(\alpha^{-1}_S -\frac{3}{8} \alpha^{-1}_{\rm em} \right)\, 
, \nonumber \\
&&\mathcal{C}_2=16 \pi\, \alpha^{-1}_{\rm em} \left(\sin^2 \theta_W -\frac{3}{8} \right) \, , \nonumber 
\end{eqnarray}
Using PDG~\cite{Agashe:2014kda} value for electroweak mixing angle $\sin^2 \theta_W = 0.23102 \mp 
0.00005$, 
strong coupling constant $\alpha_S =0.118 \pm 0.003$ and electromagnetic fine structure 
constant $\alpha_{\rm em}= 1/128.5$, determines $\mathcal{C}_1=-1996.19$, and 
$\mathcal{C}_2=-929.98$.
The other parameters can be expressed in terms of one-loop beta coefficients 
$\pmb{b}^{}$ and $\pmb{b}^{\prime}$ as,
\begin{align}
\mathcal{A}_{R}&=(8\pmb{b}_{3C}-3\pmb{b}_{2L}-5\pmb{b}_{Y})-(8\pmb{b}^{\prime}_{3C}-3\pmb{b}^{\prime
}_{2L}-3\pmb{b}^{\prime}_{2R}-2\pmb{b}^{\prime}_{BL}) \nonumber \\
\mathcal{A}_{U}&=(8\pmb{b}^{\prime}_{3C}-3\pmb{b}^{\prime}_{2L}-3\pmb{b}^{\prime}_{2R}-2\pmb{b}^{
\prime}_{BL}) \nonumber \\
\mathcal{B}_{R}&=(5\pmb{b}_{2L}-5\pmb{b}_{Y})-(5\pmb{b}^{\prime}_{2L}-3\pmb{b}^{\prime}_{2R}-2\pmb{b
}^{\prime}_{BL}) \nonumber \\
\mathcal{B}_{U}&=(5\pmb{b}^{\prime}_{2L}-3\pmb{b}^{\prime}_{2R}-2\pmb{b}^{\prime}_{BL}) 
\label{eq:AB}
\end{align}


From Eqs. \ref{gLgR:rel1}, 
\ref{gLgR:rel2} and \ref{eq:AB} we can obtain the values of the other parameters as well as the 
breaking scales which are tabulated in Table \ref{tab:table1}.
\begin{center}
\begin{table}[h!]
\begin{tabular}{|c|c|c|c|c|c|c|}
\hline
 & $\mathcal{A}_{R}$  &  $\mathcal{A}_{U}$ & $\mathcal{B}_{R}$ & $\mathcal{B}_{U}$ & $M_{R}$(GeV) & $M_{U}$(GeV) \\
\hline
Case 1 & $-18$  &  $-49$ & $-62/3$ & $-47/3$ & $2.86 \times 10^{10}$ & $3.4 \times 10^{16}$  \\
\hline
Case 2 & $-19$  &  $-48$ & $-15$ & $-64/3$ & $6.2 \times 10^{4}$ & $7.9 \times 10^{18}$ \\
\hline
\end{tabular}
\caption{Estimated values of $SU(2)_R$ breaking scale and Grand unification scale using representative set of beta cofficients.}
\label{tab:table1}
\end{table}
\end{center}

%
Turning to the determination of neutrino masses, we use
$\langle \Phi \rangle =v_1=170.572$~{GeV}, $\langle H_{L} \rangle =v_L=34.114$~{GeV}, $M_{\delta^{+}} \simeq$~{TeV}, and the two possible values $\langle H_R \rangle
\simeq 10^{10}$~{GeV} and $ 10^{4}$~{GeV} and the resulting values are displayed in Table 
\ref{tab:table2}

\begin{center}
\begin{table}[htb]
\begin{tabular}{|c|c|c|c|c|c|c|}
\hline
 $\lambda^{\prime}$  &  $\lambda^{R}$ & $Y^{\ell}$ & $M_{R}^{\text{1-loop}}$(keV) & $M_{D}$(eV) & $M_{\nu}$(eV) \\
\hline
 $10^{-2}$  &  $10^{-3}$ & $5.86 \times 10^{-11}$ & $12.67$ & $0.1$ & $ 10^{-6}$  \\
\hline
 $1$  &  $0.5$ & $5.86 \times 10^{-12}$ & $6.3$ & $0.1$ & $1.59 \times 10^{-6}$ \\
\hline
$1$  &  $0.5$ & $4.63 \times 10^{-10}$ & $1000$ & $10$ & $ 10^{-4}$ \\
\hline
$10^{-2}$  &  $10^{-3}$ & $5.86 \times 10^{-10}$ & $126.7$ & $1$ & $10^{-4}$ \\
\hline
\end{tabular}
\caption{Estimated values of physical masses for light and heavy neutrinos using derived values of $M_D$ and radiatively 
         generated $M^{1-\text{loop}}_{L,R}$ using representative set of input model parameters.}
\label{tab:table2}
\end{table}
\end{center}

\section{Neutrinoless double beta decay}
\label{sec:beta}
As discussed earlier there are no tree level Majorana masses for neutrinos and the Majorana mass terms for both left-handed and right-handed 
neutrinos are generated through radiative mechanism. With small value of Dirac neutrino mass and keV-MeV range of heavy Majorana neutrinos,  the model can accommodate a large mixing of light and heavy neutrinos. This large light-heavy neutrino mixing 
gives new physics contributions to neutrinoless double beta decay which can saturate various 
current experimental bounds. Our results with keV-MeV range of right-handed Majorana neutrino masses and large light-heavy neutrino mixing are different 
from new physics contributions arising from short distance physics due to exchange of TeV spectrum of right-handed gauge boson as well has right-handed neutrinos (for more detailed discussion on neutrinoless double beta decay 
in left-right symmetric 
models including short distance physics can be found in in refs.~\cite{Mohapatra:1980yp, Mohapatra:1981pm, Hirsch:1996qw, Tello:2010am, 
Chakrabortty:2012mh, Patra:2012ur, Awasthi:2013ff, Barry:2013xxa, Dev:2013vxa, Ge:2015yqa, 
Awasthi:2015ota,Halprin:1983ez}.

The charged current interaction Lagrangian for leptons and quarks can be read as,
\begin{align}
 {\cal L}^{\rm lep}_{CC} &=
\frac{g_L}{\sqrt{2}}\left[\sum_{\alpha=e, \mu, \tau} \overline{\ell}_{\alpha}\, \gamma^\mu P_L {\nu}_{\alpha }\, W^{-}_{L\mu} + \mbox{h.c.}\right] 
                         \notag \\
&\hspace*{2cm}+\frac{g_R}{\sqrt{2}} \left[\sum_{\alpha=e, \mu, \tau} \overline{\ell}_{\alpha}\, \gamma_\mu P_R {N}_{\alpha}\, 
W^{-}_{L\mu} + {\rm h.c.}\right]\,, \notag \\
 {\cal L}^{\rm q}_{CC} &=
\left[ \frac{g_L}{\sqrt{2}} \overline{d} \gamma^\mu P_L u W_{L\mu}^- 
+\frac{g_R}{\sqrt{2}} \overline{d}\gamma^\mu P_R u W_{R\mu}^-  + {\rm h.c.}\right]\,, \notag 
\end{align}
The flavor neutrino eigenstates $\nu_\alpha \equiv \nu_{L \alpha}$ 
and $N_\beta \equiv \nu_{R \beta}$ are related to their mass eigenstates $\nu_i$ and $N_i$ as,  
\begin{align}
&\nu_\alpha=U_{\alpha i} \nu_i + S_{\alpha i} N_i \nonumber \\ 
&N_\beta=T_{\beta i} \nu_i + V_{\beta i} N_i \nonumber 
\end{align}
where the mixing matrices $U, V, S, T$ are given by 
\begin{equation}
\left(\begin{array}{cc}
\ U & S \\
\ T & V
\end{array}\right) = \left(\begin{array}{cc}
\ 1-\frac{1}{2}R R^{\dagger} & R \\
\ -R^{\dagger} & 1-\frac{1}{2}R^{\dagger}R 
\end{array}\right) \left(\begin{array}{cc}
\ U_\nu & 0 \\
\ 0 & U_N
\end{array}\right)
\end{equation}
such that $U_\nu, U_N$ are the diagonalising matrices of light and heavy neutrino mass matrices $M_{\nu}, M_{N}$ respectively. 
Here $R=M_{D} (M_{R}^{\text{1-loop}})^{-1}$.

In the present model, there are various contributions to neutrinoless double beta decay namely 
i) due to exchange of light right-handed neutrinos via purely left-handed currents ($W_L-W_L$ mediation) or other way around, 
ii) due to exchange of keV-MeV scale right-handed neutrinos via both left-handed and right-handed currents ($W_L-W_R$ mediation), 
iii) due to mixed helicity so called $\lambda$ diagrams which involves left-right neutrino mixing through mediation of $\nu_i, N_i$ neutrinos, 
iv) due to mixed helicity $\eta$ diagrams through mediation of $\nu_i, N_i$ neutrinos involving $W_L-W_R$ gauge boson mixing as well as left-right neutrino mixing.

The half-life for a given isotope for these contributions to neutrinoless double beta decay is given by
\begin{align}
\frac{1}{T^{0\nu}_{1/2}} &= \! G_{01}  \bigg ( \lvert \mathcal{M}_\nu  \eta^L_{\nu}+\mathcal{M}^\prime_N \eta^L_N \rvert^2 + \lvert \mathcal{M}^\prime_N \eta^R_N +\mathcal{M}_\nu \eta^R_{\nu} \rvert^2 \nonumber \\
&+ \lvert \mathcal{M}^\prime_{\lambda} (\eta^{\nu}_{\lambda}+\eta^{N}_{\lambda}) + \mathcal{M}^\prime_{\eta} (\eta^{\nu}_{\eta}+\eta^N_{\eta}) \rvert^2 \bigg )
\label{eq:halflife}
\end{align}
where $G_{01}$ represents the standard $0\nu\beta\beta$ phase space factor, the $\mathcal{M}_i$  represent the nuclear matrix elements for the different exchange processes 
and $\eta_i$ are the dimensionless particle physics parameters presented in table~\ref{tableNME}. 

In the present model, we have discussed two different scenarios for gauge coupling unification predicting different values of left-right symmetry breaking scale and thereby, can result 
one-loop generated right-handed neutrinos both lighter and heavier than $100$~MeV, typical momentum exchange of the process \cite{Blennow:2010th}. It is notable that 
the relevant nuclear matrix element changes; for $M_i \gg 100$~MeV it approaches $\mathcal{M}^\prime_N \to \mathcal{M}_N$ whereas for $M_i \ll 100$~MeV it approaches 
$\mathcal{M}^\prime_N \to \mathcal{M}_\nu$ and similarly for $\mathcal{M}^\prime_{\lambda}$, $\mathcal{M}^\prime_{\eta}$. We limited our analysis to $M_i \ll 100$~MeV for 
which all the NMEs are presented in table\, \ref{tableNME}.
\begin{center}
\begin{table}[htb]
\begin{tabular}{|c|c|c|c|}
\hline
Isotope & $G_{01} \; (\text{yr}^{-1})$  &  $\mathcal{M}_\nu \equiv \mathcal{M}^\prime_N$ & $\mathcal{M}^\prime_{\lambda} \equiv \mathcal{M}^\prime_{\eta}$ \\
\hline
$ \text{Ge}-76$ & $5.77\times10^{-15}$ & $2.58-6.64$ &$1.75-3.76$  \\
$ \text{Xe}-136$ & $3.56 \times 10^{-14}$ & $1.57-3.85$ &  $1.92-2.49$ \\
\hline
\end{tabular}
\caption{Standard $0\nu\beta\beta$ phase space factor \cite{Kotila:2012zza} 
and nuclear matrix elements for the different exchange processes \cite{Pantis:1996py} used in the 
analysis}
\label{tableNME}
\end{table}
\end{center}

\subparagraph*{Left handed current effects:} 
The lepton number violating dimensionless particle physics parameter for standard $0\nu\beta\beta$ mechanism is given by,
\begin{align}
\label{eta:nu} 
\mathcal{\eta}_{\nu} \equiv \eta^L_\nu=\frac{1}{m_e}  \sum^{3}_{i=1} U^2_{ei}\, m_{i}
           = \frac{m^{L,\nu}_{\rm ee}}{m_e} \,,
\end{align}
where $m_e$ is the electron mass, $U_{ei}$ is the mixing element and $m_i$ is the light neutrino mass.  This can be translated into effective Majorana mass parameter as,
\begin{align}
\label{eq:mee-std}
m^{\nu}_{\rm ee} \equiv m^{L,\nu}_{\rm ee}
=\left| c^2_{12} c^2_{13} m_1 + s^2_{12} c^2_{13} m_2 e^{i\alpha} + s^2_{13} m_3 e^{i\beta} \right| \,,
\end{align}
where $s_{12} = \sin\theta_{12}$, $c_{12} = \cos\theta_{12}$, etc the sine and cosine of the oscillation angles. and the unconstrained Majorana phases $0 \leq \alpha,\beta < 2\pi$.

In addition, there is a new physics contribution to $0\nu\beta\beta$ mechanism due to purely left-handed current effects with the exchange of right-handed neutrinos as,
\begin{align}
\label{eta:nu} 
\mathcal{\eta}^{L}_{N} =\frac{1}{m_e}  \sum^{3}_{i=1} S^2_{ei} M_i
           = \frac{m^{L,N}_{\rm ee}}{m_e} \,. 
\end{align}
Here $S_{ei}$ is the left-right neutrino mixing whose strength depends upon the relative values of tree level Dirac neutrino mass $M_D$ and one-loop 
generated right-handed Majorana neutrino mass $M_R$ with $M_R > M_D$ and $M_i$ is the mass of right-handed neutrinos.

\subparagraph*{Right-handed current effects:}
The new physics contribution to $0\nu\beta\beta$ mechanism arising from the purely right-handed currents via the exchange of right-handed neutrinos 
yields the lepton number violating dimensionless particle physics parameter as,
\begin{align}
\label{eta:N} 
\mathcal{\eta}^{R}_N &= 
	\frac{1}{m_e}  \left(\frac{g_R}{g_L}\right)^4 \left(\frac{M_{W_L}}{M_{W_R}}\right)^4  
              \sum^{3}_{i=1} V^{*2}_{ei}M_i \,.
\end{align}
In the present scenario we have $g_L = g_R$, or else the new contributions are rescaled by the ratio between these two couplings. 
This contribution is proportional to the standard parameter $\eta_\nu$  and for $M_i \approx m_i$, the contribution becomes negligible 
because of the strong suppression from the heavy right-handed gauge boson $W_R$ mass.

Similarly, the other contribution arising from purely right-handed current effects due to exchange of light neutrinos, 
$\mathcal{\eta}^{R}_\nu= \frac{1}{m_e}  \left(\frac{g_R}{g_L}\right)^4 \left(\frac{M_{W_L}}{M_{W_R}}\right)^4   
                                             \sum^{3}_{i=1} T^{*2}_{ei}M_i \,$ is indeed negligible because of large suppression due to 
                                             the factor, $\left(\frac{M_{W_L}}{M_{W_R}}\right)^4 $.
\subparagraph*{Mixed current effects-$\lambda$ and $\eta$ diagrams:}
There are new physics contributions to $0\nu\beta\beta$ mechanism arising from the effect of both left and right handed currents are as follows, 
\begin{align} 
&\mathcal{\eta}^{\nu}_\lambda =  \left(\frac{g_R}{g_L}\right)^2 \left(\frac{M_{W_L}}{M_{W_R}}\right)^2 \sum_i U_{ei} T^*_{ei} =\frac{{\large \bf  m}_{\rm ee,\lambda}^{\nu}}{\lvert p \rvert}, \nonumber \\
&\mathcal{\eta}^{N}_\lambda=  \left(\frac{g_R}{g_L}\right)^2 \left(\frac{M_{W_L}}{M_{W_R}}\right)^2  \sum_i S_{ei} V^*_{ei} = \frac{{\large \bf  m}_{\rm ee,\lambda}^{N}}{\lvert p \rvert} \nonumber \\
&\mathcal{\eta}^{\nu}_\eta = \left(\frac{g_R}{g_L} \right) \tan{\xi} \sum_i U_{ei} T^*_{ei}=\frac{{\large \bf  m}_{\rm ee,\eta}^{\nu}}{\lvert p \rvert}, \nonumber \\
&\mathcal{\eta}^{N}_\eta = \left(\frac{g_R}{g_L} \right) \tan{\xi} \sum_i S_{ei} V^*_{ei} = \frac{{\large \bf  m}_{\rm ee,\eta}^{N}}{\lvert p \rvert}
\end{align}

In our case, all the factors of $\frac{g_R}{g_L}$ are unity.

\subsection{Numerical Results}
We intend to examine the new physics contributions which can give sizeable effects and can saturate 
the experimental limit. The translated bound on the effective Majorana mass parameter has been derived for 
various isotopes~\cite{Guzowski:2015saa,Ge:2015bfa},
\begin{eqnarray}
\label{bound}
&&\left| m^\nu_{ee} \right| \leq (0.22 - 0.53)~\mbox{eV} \quad \quad \mbox{For}~^{76}\mbox{Ge}\, \nonumber \\
&&\left| m^\nu_{ee} \right| \leq (0.36 - 0.90)~\mbox{eV} \quad \quad \mbox{For}~^{100}\mbox{Mo}\, \nonumber \\
&&\left| m^\nu_{ee} \right| \leq (0.27 - 1.00)~\mbox{eV} \quad \quad \mbox{For}~^{130}\mbox{Te}\, \nonumber \\
&&\left| m^\nu_{ee} \right| \leq (0.15 - 0.35)~\mbox{eV} \quad \quad \mbox{For}~^{136}\mbox{Xe}\,
\end{eqnarray}


One can numerically estimate the half-life for $0\nu\beta\beta$ decay of the isotope or effective Majorana mass 
parameter $m^\nu_{ee}$  (or dimensionless particle physics parameters $\eta$) using the allowed range of model 
parameters. We used phase space factors and nuclear matrix elements as displayed in Table\ref{tableNME}. The other 
model parameters are fixed as 
\begin{align}
g_R = g_L, M_{\delta^{+}} \approx 5\text{ TeV}\,, \nonumber \\ 
M_{N_i} = \text{keV}-\text{MeV}, \langle H_R \rangle \simeq \left[ 10^{10}\text{GeV}, 10^{4}\text{GeV} \right]\,.
\end{align}
Using these model parameters, the estimated effective Majorana mass parameters are presented in Table\ref{tablemee}, 
\begin{center}
\begin{table}[htb]
\begin{tabular}{|c|c|c|}
\hline
Isotope & $m_{ee}^{L,N}(\text{eV})$  &  ${\large \bf m}_{\rm ee,\lambda} \simeq {\large \bf m}_{\rm ee,\eta}$(eV)  \\
\hline
$^{76}\text{Ge}$ & $0.111-0.285$ & $39.145-84.106$  \\
\hline
$^{136}\text{Xe}$ & $0.067-0.163$ & $20.6-26.7$ \\
\hline
\end{tabular}
\caption{Numerical values of effective Majorana mass parameters due to various new physics contributions 
        within the present framework due to keV-MeV range of heavy Majorana neutrinos and sizable light-heavy 
        neutrino mixing.}
\label{tablemee}
\end{table}
\end{center}

In the analysis of gauge coupling unification discussed in the earlier section, we have considered two different scenarios predicting the left-right symmetry breaking scale as 
i) $M_R = 10^{10}$ GeV and ii) $M_R = 10^{4}$ GeV. For the case $M_R = 10^{10}$ GeV and thereby, the masses of right-handed gauge bosons $W_R,Z_R$ of the same scale, the scenario is far away from the reach of LHC. Also the ratio $\frac{M_{W_{L}}}{M_{W_{R}}}$ and  $W_L-W_R$ mixing i.e,  $\tan{\xi}$ are negligible, and thus the new physics contributions to neutrinoless double beta decay arising from purely right-handed currents and mixed current effects like $\lambda$ and $\eta$-diagrams are negligible. And due to negligle heavy-light neutrino sector mixing, $\frac{M_D}{M_{R}^{\text{1-loop}}} \sim 10^{-5}$ the contributions arising from purely left-handed currents with the exchange of light as well as heavy neutrinos are also negligible.  So, there is no new physics contribution for $0\nu\beta\beta$ for this case. As it is known the standard mechanism for $0\nu\beta\beta$ transition due to exchange of light neutrinos cannot be sensitive enough to be probed at current experiments 
for normal hierarchical (NH) and inverted hierarchical (IH) case, while the quasi-degenerate (QD) pattern is ruled out on account of the cosmology data.  
But the effective mass parameter $m_{ee}^{L,N}$, arising from purely left-handed currents and due to the exchange of heavy neutrinos, is estimated to be around 
0.1-1.0 eV while the heavy Majorana neutrinos mass $M_i$ lie in the range keV-MeV and light-heavy neutrino mixing $S_{ei}$ is around $10^{-5}$. The numerical estimation in terms of effective Majorana parameters is presented in Table.\ref{tablemee}. 

On the other hand, if we consider the $SU(2)_R$ breaking scale at about $10^4$ GeV consistent with the gauge coupling unification, the right-handed gauge bosons $Z_R,W_R$ 
are in TeV range and hence, can have rich LHC phenomenology. In addition, the new physics contributions arising from purely right-handed currents and mixed current effects 
like $\lambda$ and $\eta$-diagrams are large as the contributions are arising from large light-heavy neutrino mixing. 
In fact for the same range of input parameters, the effective mass parameter comes out as, ${\large \bf m}_{\rm ee,\lambda} \simeq 62.37$~eV. Thus, the new physics contributions are indeed large enough to saturate the 
experimental bound.

The lightest neutrino mass can also be bounded from radioactive beta decay studies, $m_\beta=\sqrt{\sum_i U^2_{ei} m_i}$ 
for which KATRIN \cite{Osipowicz:2001sq} gives the bound as $m_\beta < 0.2$~eV. From cosmology a direct limit can be placed on sum of light neutrino masses $m_\Sigma\equiv \sum_i m_i$. At present, the bound on the sum of light 
neutrino masses is $m_\Sigma < 0.23$~eV derived from Planck+WP+highL+BAO data (Planck1) at 95\% C.L. while 
$m_\Sigma < 1.08$~eV from Planck+WP+highL (Planck2) at 95\% C.L. \cite{Ade:2013zuv}. 

\begin{figure}[t!]
\includegraphics[width=0.95\linewidth]{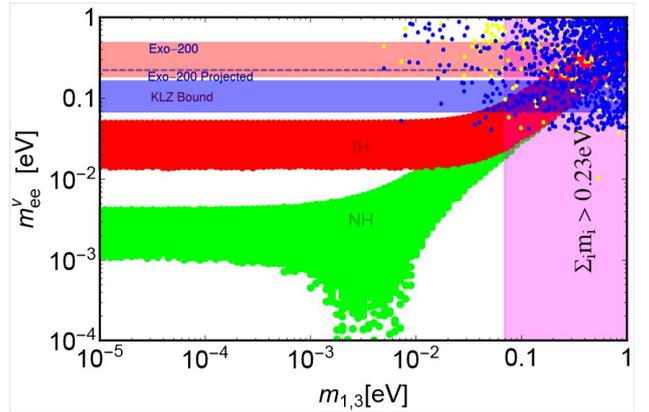}
\caption{Plots for effective Majorana mass due to standard mechanism for neutrinoless double beta decay due to exchange of light neutrinos for normal hierarchy (horizontal green shaded colour) and 
for inverted hierarchy (horizontal red shaded colour). The limit on sum of light neutrino masses from cosmological data is displayed in vertical shaded region. The other horizontal shaded coloured regions 
are displayed for bounds coming from different neutrinoless double beta decay experiments. The yellow and blue dots represent new physics contributions arising from so called $\lambda$ and $\eta$ diagrams.
}
\label{fig:mee_std}
\end{figure}
It is quite evident from Fig.\ref{fig:mee_std} that standard mechanism for neutrinoless double beta decay due to exchange of light active neutrinos are not saturated by current 
neutrinoless double beta decay experimental bound i.e KamLAND-Zen and EXO experiments for normal and inverted hiararchy. The quasi-degenerate pattern of light neutrinos 
masses are disfavoured from Cosmological bound even though they saturate the experimental bound. Thus, we need new physics contributions to confirm this if any event is found in near future.

Here we shall discuss how the present framework gives new physics contributions to neutrinoless double beta decay via so called $\lambda-$, 
$\eta-$ diagrams which are different from short distance physics mostly discussed in refs~\cite{Mohapatra:1980yp, Mohapatra:1981pm, Hirsch:1996qw, Tello:2010am, 
Chakrabortty:2012mh, Patra:2012ur, Awasthi:2013ff, Barry:2013xxa, Dev:2013vxa, Ge:2015yqa, 
Awasthi:2015ota,Halprin:1983ez}. The new physics contributions which can saturate the experimental bound are as follows
\begin{itemize}
\item due to mixed helicity so called $\lambda$ diagrams which involves left-right neutrino mixing through mediation of $\nu_i, N_i$ neutrinos, 
\item due to mixed helicity $\eta$ diagrams through mediation of $\nu_i, N_i$ neutrinos involving $W_L-W_R$ gauge boson mixing as well as left-right neutrino mixing.
\end{itemize}
The main difference is the scale of right-handed Majorana neutrino masses which is taken here in the range of keV-MeV while the usual discussions on short 
distance new physics contributions assumed TeV spectrum of $W_R$ gauge boson mass as well as GeV-TeV range right handed Majorana neutrino masses. 
Because of keV-MeV range right-handed Majorana neutrinos i.e, $M^2_i \ll |p^2|$ where $p$ is the neutrino virtuality taken to be around 100 MeV range and 
thus, yields different calculations for this rare process.

\section{Comments on Cosmological constraints}{\label{cosmo}}
In this section we want to discuss about various cosmological constraints consistent with our result regarding the keV-MeV range right-handed (RH) neutrino mass. In our case, we have studied two different left-right breaking scale $\langle H_R \rangle$ for which we have found the corresponding 1-loop generated heavy right-handed neutrino masses $M_R^{\text{1-loop}}$ presented in Table~\ref{tab:table2}. For the case, $\langle H_R \rangle \approx 10^{10}$ GeV, we can get $M_R^{\text{1-loop}} \approx 0.01$ MeV, which clearly constraints the bound from cosmology. For the case $\langle H_R \rangle \approx 10$ TeV where $M_{W_{R}}$ is lying in 1-10 TeV range, $M_R^{\text{1-loop}} \approx$ few keV. Now we want to discuss that this keV-MeV scale RH neutrinos is consistent with the big-bang nucleosynthesis (BBN) bound and from the over closing of the universe.  

A warm dark matter (DM) candidate \cite{Pagels:1981ke,Peebles:1982ib,Bond:1982uy,Olive:1981ak}, with mass lies around few keV, works as well as cold DM candidate for the large scale structure formation. Also, it can suppress the small scale structure formation via the free streaming mechanism~\cite{Colombi:1995ze}. This framework actually has been discussed as a possible solution to the problems of cuspy DM halo profiles as well as overpopulated low mass scale satellite galaxies. Previously, the idea of having RH neutrinos as warm DM candidate with mass around few keV was introduced in \cite{Olive:1981ak, Dodelson:1993je}. If the left-right breaking scale can be $\mathcal{O}$(TeV) scale, not far above the electroweak breaking scale, due to presence of gauge interaction, RH neutrino can play the role of DM having a similar relic density as one of the light neutrinos. Various previous studies explaining the connection between such keV scale RH neutrino scenario and cosmological constraints on their mass, lifetime etc. can be found in the following references~\cite{Scherrer:1984fd, Asaka:2006ek,Bezrukov:2009th,Nemevsek:2012as,Nemevsek:2012cd}. From this discussion, we can easily point out that our analysis with keV-MeV range RH neutrino can be visualised as such warm DM candidate which can satisfy the various cosmological and astrophysical bounds summarised in the table \ref{astro}\cite{Nemevsek:2012cd},
\begin{table}[htb]
\begin{tabular}{|c|c|c|c|}
\hline
Constraints & $M_R^{1-loop}$ & $\tau_N$ & $M_{W_R}$ \\
\hline
Dwarf Galaxy & $\gtrsim 0.4-0.5$~keV & $-$ & $-$ \\
\hline
Lyman-$\alpha$ & $\gtrsim 0.5-1$~keV & $-$ & $-$ \\
\hline 
BBN and CMB & $-$ & $\lesssim 1.5$~sec & $-$ \\
\hline
$0\nu\beta\beta$ & $-$ & $-$ & $\gtrsim 6-8$~TeV \\
\hline
\end{tabular} 
\caption{Various bounds from astrophysical and cosmological signatures within the LRSM framework}
\label{astro}
\end{table}  

From the cosmological studies, we can infer that over abundance of keV RH neutrinos due to the fact that at high temperature the $SU(2)_R$ gauge interaction keep them in thermal equilibrium, when the right-handed gauge boson mass $M_{W_R}$ lies around TeV scale. 

The only way out of the over-abundance of keV neutrinos is to dilute the number density of lightmost right-handed neutrino $N_1$ by the so-called entropy production mechanism due to the late entropy of some massive particles which dominate the universe. Such late decay should involve some relativistic light Standard model particles that quickly equilibriated with the thermal plasma and also reheat the universe. As a consequence, the number density of the DM candidate is effectively reduced. In order to dilution mechanism to work, the temperature of $N_1$ should not increase, so $N_1$ itself cannot be a decay product of the other heavy decaying RH neutrinos $N_{2,3}$ which actually play the role of diluters in our scenario. In order to achieve a sizable dilution, the mass of these diluters $m_{N_{2,3}}$ should not exceed its freeze-out temperature $T_f$~\cite{Nemevsek:2012cd}. 

As the universe cools down, the total energy density of the universe can be temporarily dominated by sufficiently heavy and long-lived RH neutrinos. After RH neutrino decays, the energy
density is transferred into that of radiation. In the sudden decay approximation, the relation between t $\approx \tau_N$  and reheating temperature of the universe $T_r$ can be approximated as~\cite{Scherrer:1984fd},
\begin{equation}
T_r \simeq 1.22 ~\text{MeV}~ \left(\frac{1~ \text{sec}}{\tau_{N_i}}\right)^{1/2}
\label{reheat}
\end{equation}
where $i=2, 3.$

In order to begin BBN with correct proton-neutron number ratio, we need $T_r \geqslant$ some MeV, so from eqn.~\ref{reheat}, we can say that $\tau_N \lesssim \mathcal{O}$ (sec). 

Depending on the diluters' mass, $N_{2,3}$ decay (via $W_R$ boson mediation) either into a lepton + two light quarks with a lifetime,
\begin{equation}
\tau (N_{2,3} \rightarrow ljj) = 1~\text{sec}~\left(\frac{2~\text{GeV}}{m_{N_{2,3}}}\right)^5 \left(\frac{M_{W_R}}{100~\text{TeV}}\right)^4
\end{equation}
or if $m_{N_{2,3}} \gtrsim m_{\pi}+m_{l}$ then into a lepton + a pion with a lifetime,
\begin{equation}
\tau (N_{2,3} \rightarrow l\pi) = 1~\text{sec}~\left(\frac{250~\text{MeV}}{m_{N_{2,3}}}\right)^3 \left(\frac{M_{W_R}}{5~\text{TeV}}\right)^4 \left(\frac{0.002}{g(x_l,x_{\pi})}\right)
\end{equation}
with $g(x_l, x_{\pi})=[(1-x_l^2)^2-x_{\pi}^2(1+x_l^2)][(1-(x_l+x_{\pi})^2)(1-(x_{\pi}-x_l)^2)]^{1/2}$ where $x_{\pi ,l}=\frac{m_{\pi ,l}}{m_{N_{2,3}}}$. Here, the produced lepton $l$ can be either $\mu$ or $e$, detailed discussion can be found in \cite{Nemevsek:2012cd}. To have a rich collider phenomenology with $M_{W_R} \sim 10$ TeV, the dilution process $N_{2,3} \rightarrow l\pi$ should be the dominant one.

Now we can turn our attention to discuss the stability of remaining keV scale RH neutrino $N_1$ as viable warm dark matter candidate~\cite{Chakraborty:2014tma,Patra:2014pga}. In our analysis, we have for both the left-right breaking scale scenarios, the RH and LH neutrino sector have very small mixing i.e, $\frac{M_D}{M_R^{\text{1-loop}}} \leqslant 10^{-5}$, which actually forbids the LH neutrino oscillation back to the RH neutrinos (which again creates the overabundance of keV scale RH neutrino dark matter). But, this tiny mixing causes a right handed neutrino ($N_1$) to decay into a light active neutrino and a monochromatic photon line of energy $E_{\gamma} = m_{N_1} /2$. The decay width of such two-body radiative decay process (via mediation of $W$-boson) can be formulated~\cite{Patra:2014pga} as
\begin{equation}
\Gamma_{N_1 \rightarrow \nu \gamma} = \frac{9\alpha G^2_F}{1024 \pi^2}sin^2 2\theta~ m^5_{N_1}
\end{equation}

where, $\alpha$ and $G_F$ are the electromagnetic fine-structure constant and universal Fermi coupling constant respectively. Also, 3-body decay process via neutral $Z$-boson mediation is also permissible with the corresponding decay width~\cite{Patra:2014pga},
\begin{equation}
\Gamma_{N_1 \rightarrow 3\nu} = \frac{4G_F^2}{384\pi^2}sin^2 2\theta~ m^5_{N_1}
\end{equation}

From observed 3.5 keV X-ray line signal and the results from various recent studies, we can decude the mass and mixing of a RH neutrino should be indeed very small to make itseld a viable DM candidate which survives much longer than the universe. With the corresponding $m_{N_1} \approx $ few keV and $sin^2(2\theta) \sim 10^{-10}$, we can have the decay lifetime to be~\cite{Patra:2014pga},
\begin{equation}
\tau_{N_1 \rightarrow 3\nu_L} \approx 10^{28} \left(\frac{10^{-10}}{sin^2(2\theta)}\right)\left(\frac{\text{keV}}{m_{N_1}}\right)^5 \text{sec}
\end{equation}
\begin{equation}
\tau_{N_1 \rightarrow \nu_L \gamma} \approx 10^{30} \left(\frac{10^{-10}}{sin^2(2\theta)}\right)\left(\frac{\text{keV}}{m_{N_1}}\right)^5 \text{sec}
\end{equation}
So, assuming a RH neutrino DM with the mass at around few keV and its very small mixing with LH neutrino sector can easily satisfy the stability criteria. Some other experiments on X-ray signal with different energy ($\sim$ few keV) put a constraint on the mixing between the RH and LH sector~\cite{ASmirnov:2006bu,Boyarsky:2009ix,Abazajian:2001nj} as, 
\begin{equation}
\theta^2 \leqslant 1.2 \times 10^{-5} \left(\frac{\text{keV}}{m_{N_1}}\right)^5
\end{equation}  
Also, another necessary point of a DM model is to predict exact relic density of DM in today's universe. Relic abundance of keV mass RH neutrino DM can be supported some specific production mechanism which involve very intricate calculations. Among them Dodelson-Widrow mechanism~\cite{Dodelson:1993je} gives us,
\begin{equation}
\Omega_{N_1}h^2 \approx 0.3 \left(\frac{sin^2 (2\theta)}{10^{-10}}\right)\left(\frac{m_{N_1}}{100~ \text{keV}}\right)^2
\end{equation}   
which clearly shows that our $N_1$ neutrinos with mass of $\mathcal{O}(10^2)$~keV and light-heavy neutrino mixing of $\mathcal{O}(10^{-5})$, we can have correct DM relic density of universe.

\section{Left-right symmetry without scalar bidoublet and Universal Seesaw}
\label{sec:lrsm-universal}
In left-right symmetric models with only Higgs doublets with $B-L=1$ and without scalar bidoublet, there are no Dirac masses for 
quarks and leptons. In order to have Dirac masses for quarks and charged leptons, we add vector like fermions presented in Tab.~\ref{tab:LR2}. The motivation for inclusion of vector like fermions is to provide non-zero masses to quarks and leptons through a common 
seesaw called as universal seesaw~\cite{Davidson:1987mh,Rajpoot:1986nv,Chang:1986bp,Babu:1988yq,Babu:1988mw,Babu:1989rb,Balakrishna:1987qd,Balakrishna:1988bn,Balakrishna:1988ks},
\begin{gather}
	U_{L,R} \sim (\mathbf{1},\mathbf{1},\mathbf{4/3},\mathbf{3})\,,\quad D_{L,R} \sim (\mathbf{1},\mathbf{1},\mathbf{-2/3},\mathbf{3})\, , \nonumber \\
	E_{L,R}\sim (\mathbf{1},\mathbf{1},\mathbf{-2},\mathbf{1}).
\end{gather}


%
\begin{table}[h!]
\begin{center}
\begin{tabular}{|c|c|c||c|c|}
\hline
Field     & $ SU(2)_L$ & $SU(2)_R$ & $B-L$ & $SU(3)_C$ \\
\hline
$q_L$     &  2         & 1         & 1/3   & 3   \\
$q_R$     &  1         & 2         & 1/3   & 3   \\
$\ell_L$  &  2         & 1         & -1    & 1   \\
$\ell_R$  &  1         & 2         & -1    & 1   \\
\hline
$U_{L,R}$ &  1         & 1         & 4/3   & 3   \\
$D_{L,R}$ &  1         & 1         & -2/3  & 3   \\
$E_{L,R}$ &  1         & 1         & -2    & 1   \\
\hline
 $H_L$    &  2         & 1         & 1    & 1   \\
 $H_R$    &  1         & 2         & 1    & 1   \\
\hline
\end{tabular}
\end{center}
\caption{LRSM representations of extended field content for universal seesaw.}
\label{tab:LR2}
\end{table}

The extension of left-right symmetric models with isosinglet vector-like copies of fermions with additional 
neutral vector like fermions can be found in refs~~\cite{Mohapatra:2014qva,Dev:2015vjd,Deppisch:2016scs,Gabrielli:2016vbb,Patra:2012ur,Hati:2018tge,Deppisch:2017vne}. 
and their embedding in gauged flavour groups 
with left-right symmetry \cite{Guadagnoli:2011id} or quark-lepton symmetric models \cite{Joshi:1991yn}. However, we would like to 
discuss here the possibilities of lepton number violation with the inclusion of singly charged scalar. 

The relevant Yukawa part of the Lagrangian is given by
\begin{align}
\mathcal{L} = 
	&- \sum_{X} ( \lambda_{SXX} S \overline{X} X + M_X \overline{X} X )  \nonumber\\
	&- (\lambda_U^L \tilde{H}_L \overline{q}_L U_R 
     + \lambda_U^R \tilde{H}_R \overline{q}_R U_L \nonumber\\
  &+\phantom{(}\lambda_D^L H_L \overline{q}_L D_R
	   + \lambda_D^R H_R \overline{q}_R D_L \nonumber\\
  &+\phantom{(}\lambda_E^L H_L\overline{\ell}_L E_R 
	   + \lambda_E^R H_R\overline{\ell}_R E_L \nonumber\\
  &+\lambda^L \ell^T_L C \ell_L \delta^+ + \lambda^R \ell^T_R C \ell_R \, \delta^+ + \mbox{h.c.}\,
\label{2.1}
\end{align}
where $X = U, D, E$, $\tilde{H}_{L,R}$ denotes $\tau_2 H_{L,R}^\ast$, where $\tau_2$ is the usual second Pauli matrix. 

After spontaneous symmetry breaking, the Dirac masses for quarks and leptons are given by,
\begin{align}
\label{2.3}
	M_{uU}    = \begin{pmatrix} 0 & \lambda_U^L v_L \\ \lambda_U^R v_R & M_U \end{pmatrix}, \,
	M_{dD}    = \begin{pmatrix} 0 & \lambda_D^L v_L \\ \lambda_D^R v_R & M_D \end{pmatrix}, 
	\nonumber\\
	M_{e E}   = \begin{pmatrix} 0 & \lambda_E^L v_L \\ \lambda_E^R v_R & M_E \end{pmatrix}, \,
	M_{\nu N} = \begin{pmatrix} 0 & \lambda_N^L v_L \\ \lambda_N^R v_R & M_N \end{pmatrix},
\end{align}

Here Dirac neutrino masses are generated at two loop level with the $W_L-W_R$ mixing derived at one-loop and found to be suppressed i.e, less than 
$0.01$~eV. However, there is no lepton number violation in this set up. We discuss below another framework to have lepton number violation in left-right theories without scalar bidoublet.

We accomodate lepton number violation by extending scalar sector consisting of $SU(2)_{L,R}$ doublets and triplets, but the conventional scalar bidoublet is absent. At first stage, the left-right symmetric model is broken down to SM by $H_R \equiv (h_R^0, h^-_R)^T \equiv [1,2,-1,1]$ and subsequently, SM to low energy theory is acheived by $H_L\equiv (h_L^0, h^-_L)^T \equiv [2,1,-1,1]$ with following VEV structure, 
\begin{align}
	\langle H_R \rangle = \begin{pmatrix} \frac{v_R}{\sqrt{2}} \\ 0 \end{pmatrix}, \quad 
	\langle H_L \rangle = \begin{pmatrix} \frac{v_L}{\sqrt{2}} \\ 0 \end{pmatrix}. 
\end{align}
Scalar triplets $\Delta_{L}$ and $\Delta_{R}$ do not get any VEV at tree level and these VEVs can induced by trilinear terms derived from scalar potential. The matrix structure for these fields 
\begin{align}
	\Delta_{L,R} &= \begin{pmatrix} \delta_{L,R}^+/\sqrt{2} & \delta_{L,R}^{++} \\ \delta_{L,R}^0 & -\delta_{L,R}^+/\sqrt{2} \end{pmatrix}\,,
\end{align}
which transform as $\Delta_L \equiv [3,1,2,1]$ and $\Delta_R \equiv [1,3,2,1]$, respectively. The particle content of the left-right models with universal seesaw which can accomodate large lepton number violating is presented in Tab.~\ref{tab:LR3}. The scalar potential of the model is given by 
 \begin{align}
&\hspace*{-0.6cm}\mathcal{V}\left(H_L, H_R, \Delta_L, \Delta_R \right) =- \mu^2_1 (H^\dagger_L H_L) - \mu^2_2 (H^\dagger_R H_R)  \nonumber \\
&  +\lambda_1  (H^\dagger_L H_L)^2 + \lambda_2  (H^\dagger_R H_R)^2+\beta_1 (H^\dagger_L H_L)  (H^\dagger_R H_R) \nonumber\\
&+ \mu^2_3  \mbox{Tr}(\Delta^\dagger_L \Delta_L) + \mu^2_4  \mbox{Tr}(\Delta^\dagger_R \Delta_R) \nonumber \\
 &+ \lambda_3  \mbox{Tr}(\Delta^\dagger_L \Delta_L)^2 + \lambda_4  \mbox{Tr}(\Delta^\dagger_R \Delta_R)^2 \nonumber \\
 &+ \beta_2 \mbox{Tr}(\Delta^\dagger_L \Delta_L)  \mbox{Tr}(\Delta^\dagger_R \Delta_R) \nonumber \\
&+\rho_1 ( \mbox{Tr}(\Delta^\dagger_L \Delta_L)  (H^\dagger_R H_R)+ \mbox{Tr}(\Delta^\dagger_R \Delta_R)  (H^\dagger_L H_L)) \nonumber \\
&+\rho_{2} \left(  \mbox{Tr}(\Delta^\dagger_L \Delta_L)  (H^\dagger_L H_L)+ \mbox{Tr}(\Delta^\dagger_R\Delta_R)  (H^\dagger_R H_R) \right) \nonumber \\
& +\rho_{3}\left( H_L^\dagger \Delta_L^\dagger \Delta_L H_L +H_R^\dagger \Delta_R^\dagger \Delta_R H_R\right) \nonumber \\
&+ \mu_{} \left( H^T_L i\sigma_2 \Delta_L H_L+H^T_R i\sigma_2 \Delta_R H_R \right) +\mbox{h.c.} \cdots \,.
\label{eq:scalar_lagrangian}
\end{align}
It is important to note here that the sign of $\mu^2_{1,2}$ is negative while sign of $\mu^2_{3,4}$ is positive. The minimisation condition allows non-zero VEV for Higgs doublets while there are no VEVs for scalar triplets at tree level. However, non-zero VEVs for scalar triplets are induced after Higgs doublets take non-zero VEVs and derived from trilinear coupling $\mu_{} \left( H^T_L i\sigma_2 \Delta_L H_L+H^T_R i\sigma_2 \Delta_R H_R \right)$. The idea is to break left-right symmetry with Higgs doublets and induce small VEVs for scalar triplets such that we can get light right-handed neutrino masses and their Implications to neutrinoless double beta decay. Thus, after spontaneous symmetry breaking, the VEVs for Higgs doublets and induced VEVs for scalar triplets are presented below 
\begin{align}
       & \langle H^0_L \rangle \equiv v_L/\sqrt{2}, \quad \langle H^0_R \rangle \equiv v_R/\sqrt{2},  \nonumber \\
       &\langle \Delta^0_L \rangle \equiv u_L/\sqrt{2}\,, \quad   \langle \Delta^0_R \rangle \equiv u_R/\sqrt{2} \,.
\end{align}
the scalar potential read as,
 \begin{align}
&\hspace*{-0.6cm}\mathcal{V}\left(\langle H_L \rangle , \langle H^0_R \rangle , \langle \Delta^0_L \rangle , \langle \Delta^0_R \rangle \right) =  \nonumber \\
&-\frac{1}{2} \mu^2_1 v^2_L - \frac{1}{2} \mu^2_2 v^2_R +\frac{1}{2} \mu^2_3 u^2_L + \frac{1}{2} \mu^2_4 u^2_R \nonumber \\
&+\frac{1}{4} \lambda_1 v^4_L + \frac{1}{4} \lambda_2 v^4_R +\frac{1}{4} \lambda_3 u^4_L + \frac{1}{4} \lambda_4 u^4_R \nonumber \\
&+\frac{1}{4}  \beta_1 v^2_L v^2_R +\frac{1}{4}  \beta_2 u^2_L u^2_R - \frac{1}{2\sqrt{2}} \mu \left(v^2_L u_L + v^2_R u_R \right)  \nonumber \\
&+\frac{1}{4} \rho_1 \left(v^2_R u^2_L +  v^2_L u_R^2\right) +\frac{1}{4} \rho_2 \left(v^2_L u^2_L +  v^2_R u_R^2\right) + \cdots 
\label{eq:scalar_vev}
\end{align}

The relation between VEVs of scalar triplets and Higgs doublets are as follows,
\begin{align}
	 u_L = \frac{\mu v_L^2}{M^2_{\delta_L^0}}\,, \quad
	 u_R = \frac{\mu v_R^2}{M^2_{\delta_R^0}}\,.
\end{align}

The Majorana masses for light and heavy neutrinos by induced VEV of scalar triplets are
$\langle \Delta_L \rangle =u_L$ and $\langle \Delta_R \rangle =u_R$ giving the neutral lepton mass matrix in the basis $(\nu_L, \nu_R)$ given by
\begin{align}
	M_\nu= 
	\left(\begin{array}{cc}
		f u_L  & 0  \\
   	    0 & f u_R
\end{array} \right) \,.
\label{eqn:numatrix}       
\end{align}

The physical masses for light neutrinos are given by $m_\nu=f u_L$ and 
for heavy neutrinos as $M_R=f u_R$. These heavy neutrinos and scalar triplets can mediate large lepton number violation and give new physics contributions to neutrinoless double beta dceay. 

\begin{table}[h]
\begin{center}
\begin{tabular}{|c|c|c||c|c|}
\hline
Field     & $ SU(2)_L$ & $SU(2)_R$ & $B-L$ & $SU(3)_C$ \\
\hline
$Q_L$     &  2         & 1         & 1/3   & 3   \\
$Q_R$     &  1         & 2         & 1/3   & 3   \\
$\ell_L$  &  2         & 1         & -1    & 1   \\
$\ell_R$  &  1         & 2         & -1    & 1   \\
\hline
$U_{L,R}$ &  1         & 1         & 4/3   & 3   \\
$D_{L,R}$ &  1         & 1         & -2/3  & 3   \\
$E_{L,R}$ &  1         & 1         & -2    & 1   \\
\hline
 $H_L$    &  2         & 1         & -1    & 1   \\
 $H_R$    &  1         & 2         & -1    & 1   \\
 $\Delta_L$      &  3         & 1         & 2     & 1   \\
  $\Delta_R$      &  1         & 3         & 2     & 1   \\
\hline
\end{tabular}
\end{center}
\caption{Field content of the LRSM with universal seesaw.}
\label{tab:LR3}
\end{table}

\section{Conclusions}{\label{sec5}}
We have considered a version of left-right symmetric model giving rise to Majorana masses for left-handed and 
right-handed neutrinos through a radiative mechanism in turn contributing to neutrinoless double beta decay. 
The radiative mechanism for Majorana masses is achieved through the introduction of the charged scalar $\delta^+(1_L, 1_R, 2_{BL}, 1_C)$. The light neutrino mass generation is explained via the type-I seesaw 
mechanism with keV scale for right-handed neutrinos and few eV scale for Dirac neutrino mass using 
suppressed value of Yukawa coupling as in the Table~\ref{tab:table2}. This choice of model 
parameters can 
saturate the experimental bound of GERDA and KamLAND-Zen experiments on neutrinoless double beta decay.

We embedded this model within a non-SUSY $SO(10)$ GUT framework with successful gauge coupling unification. 
The simplest possibility gives rise to unification at $10^{16}$~GeV with the scale of left-right symmetry 
breaking around $10^{10}$~GeV. Alternatively, an extension of the framework with addition of scalar species permits the intermediate left-right 
symmetry breaking at TeV scale so that the right-handed gauge bosons $Z_R, W_R$ can have interesting Collider 
as well as low energy phenomenology signatures. The two possible values of left-right symmetry breaking scale permit the keV to MeV range for Majorana masses for the right handed neutrinos, in turn leading to sizeable new contributions to neutrinoless 
double beta decay.

Also, we then briefly explained the viability of our model through the one-loop level generated heavy right-handed neutrino masses $M_R^{\text{1-loop}}$ (lies in the range of keV-MeV) which clearly saturate various constarints obtained from astrophysical, cosmological as well as terrestrial experiments within the framework of LRSM. 

We also discussed left-right symmetric models with Higgs doublets $H_{L,R}$ and singly charged scalar $\delta^{+}$ without having scalar bidoublet. 
In the absence of scalar bidoublet, one can not write down the Dirac masses for all fermions including quarks and leptons at tree level. With the inclusion 
of additional vector like fermions, all the fermions get their masses through universal seesaw. 
\appendix
\section{Lagrangian for this Left-Right Theories with lepton number violation}
\label{sec:a1}
The Lagrangian for this left-right symmetric model (omitting the $SU(3)_C$ structure for simplicity) is given by
\begin{eqnarray}
\mathcal{L}_{\rm LR}&=& \mathcal{L}^{\rm }_{\rm scalar} + \mathcal{L}^{\rm gauge}_{\rm Kin.} 
      + \mathcal{L}^{\rm fermion}_{\rm Kin.} + \mathcal{L}_{\rm Yuk}
\end{eqnarray}
where the individual parts can be written as
\begin{widetext}
\begin{eqnarray}
\mathcal{L}^{\rm }_{\rm scalar} &=&
      \mbox{Tr}\big[\left(\mathcal{D}_\mu \Phi\right)^\dagger \left(\mathcal{D}^\mu \Phi\right) \big] 
      +\left(\mathcal{D}_\mu \delta^+\right)^\dagger \left(\mathcal{D}^\mu \delta^+\right)
      + \left(\mathcal{D}_\mu H_L\right)^\dagger \left(\mathcal{D}^\mu H_L\right) \nonumber \\
      &&+ \left(\mathcal{D}_\mu H_R\right)^\dagger \left(\mathcal{D}^\mu H_R\right) 
       - \mathcal{V}(\Phi, H_L, H_R, \delta^+) 
\end{eqnarray} 
Defining $\Phi \equiv \Phi_1$ and $\Phi_2=\tau_2 \Phi^* \tau_2$, the scalar potential can be written as follows
\begin{eqnarray}     
\mathcal{V}(\Phi, H_L, H_R, \delta^+)&=&
-\sum_{i,j=1,2} \frac{\mu_{\phi ij}^{2}}{2}~\mbox{Tr}(\Phi_{i}^{\dagger}\Phi_{j}) 
+\sum_{i,j,k,l=1,2}\frac{\lambda_{\phi ijkl}}{4}~\mbox{Tr}(\Phi_{i}^{\dagger}\Phi_{j})
~\mbox{Tr}(\Phi_{k}^{\dagger}\Phi_{l}) 
 \nonumber \\
   &&+\sum_{i,j,k,l=1,2}\frac{\Lambda_{\phi ijkl}}{4}~\mbox{Tr}(\Phi_{i}^{\dagger}\Phi_{j}\Phi_{k}^{\dagger}\Phi_{l}) 
   -\mu^2_H\, \left(H^\dagger_L H_L +H^\dagger_R H_R \right) +\lambda_1\, \left[\left(H^\dagger_L H_L\right)^2+ \left(H^\dagger_R H_R \right)^2 \right]
    \nonumber \\
   && +\lambda_2\, \left(H^\dagger_L H_L\right) \left(H^\dagger_R H_R \right)
   +\sum_{i,j} \beta_{ij}~\left(H^\dagger_L H_L +H^\dagger_R H_R \right)\mbox{Tr}(\Phi_{i}^{\dagger}\Phi_{j})  
   \nonumber \\
   &&+\sum_{i,j} \varrho_{ij}\left[H^\dagger_L \Phi_i \Phi^\dagger_j H_L+H^\dagger_R \Phi^\dagger_i \Phi_i H_R\right]
    \nonumber \\
   &&-\mu^2_\delta \delta^+ \delta^- + \lambda_\delta \left(\delta^+ \delta^- \right)^2    
   +\lambda_{\delta H}\,\left(\delta^+ \delta^-\right) \left(H^\dagger_L H_L +H^\dagger_R H_R \right)
   \nonumber \\
&&+\sum_{i,j}\lambda_{\delta \Phi ij}\,\,\left(\delta^+ \delta^-\right) \mbox{Tr}(\Phi_{i}^{\dagger}\Phi_{j})
   +\lambda^\prime H^\dagger \Phi H^*_R \delta^+\, .
   \label{sp}
\end{eqnarray}

The kinetic terms for gauge bosons is given by
\begin{eqnarray} 
\mathcal{L}^{\rm gauge}_{\rm Kin.}&=&
     -\frac{1}{4}W_{\mu\nu L}.W^{\mu\nu L}-\frac{1}{4}W_{\mu\nu R}.W^{\mu\nu R}     -\frac{1}{4}B_{\mu\nu}B^{\mu\nu} 
\end{eqnarray}
while for fermions,
\begin{eqnarray} 
\mathcal{L}^{\rm fermion}_{\rm Kin.}&=&
       i  \overline{q_{L}}\gamma^{\mu} \mathcal{D}_\mu q_{L}
      +i  \overline{q_{R}}\gamma^{\mu} \mathcal{D}_\mu q_{R}  +i  \overline{\ell_{L}}\gamma^{\mu} \mathcal{D}_\mu \ell_{L}
      +i  \overline{\ell_{R}}\gamma^{\mu} \mathcal{D}_\mu \ell_{R}
      \nonumber \\
\end{eqnarray}
where the respective covariant derivatives, in general, are as
\begin{eqnarray} 
\mathcal{D}^f_\mu=\partial_{\mu}-i\,g_L \tau^a W^a_{\mu L} -i\,g_R \tau^a W^a_{\mu R}
                  -i\,g_{BL} \frac{B-L}{2} B_{\mu} \nonumber
\end{eqnarray}
The Yukawa interaction Lagrangian can be read as
\begin{eqnarray}
\mathcal{L}_{\rm Yuk}&=&
     Y_q\, \overline{q_{L}} \Phi q_{R} + \widetilde{Y_q} \overline{q_{L}}  \widetilde{\Phi}\, q_{R}     +Y_\ell\, \overline{\ell_{L}} \Phi \ell_{R} + \widetilde{Y_\ell} \overline{\ell_{L}} \widetilde{\Phi}\, \ell_{R}+\lambda_L \ell^T_L C \ell_L \delta^+ + \lambda_R \ell^T_R C \ell_R \, \delta^+ + \mbox{h.c.}\,
\end{eqnarray}
\section{Scalar potential minimization}
\label{sec:a2}
Putting the VEVs as given in Eq.~\ref{vev} in the scalar potential~\ref{sp} (in the limit, $v_2 \rightarrow 0$)
 the parametrized potential can be obtained as,
\begin{align}
\mathcal{V}(v_1, v_{L}, v_{R}) & = -\mu_{\phi}^{2}|v_1|^{2}+\lambda_{\phi}|v_1|^{4}+ \frac{\Lambda_{\phi}}{2} |v_1|^4 -\mu_{H}^{2}(v_{L}^{2}+v_{R}^{2})+ \lambda_{1}(v_{L}^{4}+v_{R}^{4}) \nonumber \\
&+\lambda_{2}v_{L}^{2}v_{R}^{2} +(2\beta + \varrho)|v_1|^{2}(v_{L}^{2}+v_{R}^{2})
\label{para}
\end{align}
Minimizing the Eq.\ref{para} with respect to $v_{L}$ and $v_{R}$ we get,
\begin{equation}
\frac{\partial \mathcal{V}}{\partial v_{L}}=\mu_{L}^{2}v_{L}- 2\lambda_{1}v_{L}^{3}-\lambda_{2}v_{L}v_{R}^{2} =0
\label{b}
\end{equation} 
\begin{equation}
\frac{\partial \mathcal{V}}{\partial v_{R}}=\mu_{R}^{2}v_{R}-2\lambda_{1}v_{R}^{3}-\lambda_{2}v_{R}v_{L}^{2}=0
\label{c}
\end{equation}

where 
$\mu_{L}^{2}=\mu_{R}^{2} \equiv \mu_{H}^{2}-(2\beta + \varrho)|v_1|^{2}$.\\
From Eq.\ref{b} and Eq.\ref{c}, considering $v_{L} \neq v_{R}$, we get,
\begin{equation}
v_{L}=\frac{2\lambda_{1}(v_{L}^{2}+v_{R}^{2})- \mu_{L}^{2} }{(\lambda_{2}-2\lambda_{1})v_{R}}
\end{equation}
Also we can express $|v_1|^{2}$ in terms of $v_L$ and $v_R$ as 
\begin{equation}
|v_1|^{2} = \frac{\mu_{H}^{2}+ (\lambda_{2}-2\lambda_{1})v_L v_{R}- 2\lambda_{1}(v_{L}^{2}+v_{R}^{2})}{(2\beta + \varrho)}
\end{equation}
\end{widetext}

\providecommand{\href}[2]{#2}
\begingroup
\raggedright
\endgroup

\end{document}